\documentclass[reprint,
 amsmath,amssymb,
 aps,
]{revtex4-1}

\usepackage{graphicx}% Include figure files
\usepackage{dcolumn}% Align table columns on decimal point
\usepackage{bm}% bold math
\usepackage{braket}
\usepackage{hyperref}% add hypertext capabilities
\usepackage{cleveref}
\usepackage{verbatim}
\DeclareMathOperator{\Tr}{Tr}
\usepackage{xcolor}
\def\red{\textcolor{red}}

\begin{document}

\preprint{APS/123-QED}

\title{Unconventional scaling at non-Hermitian critical points}% Force line breaks with \\

\author{R. Arouca$^{1,2}$}\thanks{r.aroucadealbuquerque@uu.nl}
\author{C. H. Lee$^{3}$}\thanks{phylch@nus.edu.sg}\author{C. Morais Smith$^{1}$}\thanks{C.deMoraisSmith@uu.nl}
\affiliation{$^1$Institute for Theoretical Physics, Center for Extreme Matter and Emergent Phenomena, Utrecht University, Princetonplein 5, 3584 CC Utrecht, The Netherlands}%Lines break automatically or can be forced with \\
\affiliation{$^2$Instituto de F\' isica, Universidade Federal do Rio de Janeiro, C.P. 68528, Rio de Janeiro, RJ, 21941-972, Brazil}%\\
\affiliation{$^3$Departament of Physics, National University of Singapore, Singapore 117542}%Lines break automatically or can be forced with 

\date{\today}% It is always \today, today,

\begin{abstract}
Critical phase transitions contain a variety of deep and universal physics, and are intimately tied to thermodynamic quantities through scaling relations. Yet, these notions are challenged in the context of non-Hermiticity, where spatial or temporal divergences render the thermodynamic limit ill-defined. In this work, we show that a thermodynamic grand potential can still be defined in pseudo-Hermitian Hamiltonians, and can be used to characterize aspects of criticality unique to non-Hermitian systems. Using the non-Hermitian Su-Schrieffer-Heeger (SSH) model as a paradigmatic example, we demonstrate the fractional order of topological phase transitions in the complex energy plane. These fractional orders add up to the integer order expected of a Hermitian phase transition when the model is doubled and Hermitianized.  More spectacularly, gap preserving highly degenerate critical points known as non-Bloch band collapses possess fractional order that are not constrained by conventional scaling relations, testimony to the emergent extra length scale from the skin mode accumulation. Our work showcases that a thermodynamic approach can prove fruitful in revealing unconventional properties of non-Hermitian critical points.
\end{abstract}

\maketitle

%\tableofcontents

\section{Introduction}
\label{sec_intro}

	Critical phase transitions have mesmerized generations of physicists, with their profound implications in conformal and statistical field theory~\cite{aizenman1987phase,zamolodchikov1989exact,cardy1992critical,dziarmaga2005dynamics,gruzberg2006stochastic,stevenson2011domain, mussardo2010statistical}, entanglement entropy ~\cite{vidal2003entanglement,korepin2004universality,ryu2006aspects,laflorencie2006boundary,swingle2010entanglement,lee2015exact,swingle2016area, chang2019ent} and thermodynamics~\cite{mussardo2010statistical, continentino2017quantum, *continentino1994quantum, sachdev_2011, vojta2003quantum, salinas2001introduction, huang1963statistical, herbut2007modern}. Placed in a non-Hermitian context, critical transitions become even more intriguing, involving the closure of various line or point gaps and exceptional points~\cite{berry2004EP, bergholtz2019exceptional, ghatak2019new,heiss2012EP,xu2016topological,hodaei2017enhanced,wang2019arbitrary,miri2019exceptional,okugawa2019exceptional,budich2019symmetry,yoshida2019ER,yoshida2019ER_2,kawabata2019classification,yang2019fermion,li2020topological,denner2020exceptional}. With the non-Hermitian skin effect (NHSE)~\cite{kunst2018biorthogonal,yao2018edge,yokomizo2019non,lee2019anatomy,lee2018tidal,song2019non,okuma2019topological,PhysRevLett.124.086801,yang2019auxiliary,lee2019unraveling,yi2020non,ma2020quantum, helbig2020generalized}, the emergent non-locality from asymmetric hoppings can allow topological transitions to occur even without gap closure~\cite{lee2019unraveling}, and the system size itself can act as a parameter that drives the model into different phases~\cite{li2020critica,liu2020helycal,lee2020ultrafast}.

	Naively, the concept of thermodynamics appears to be incompatible with non-Hermiticity, since states with complex eigenenergies cannot be in equilibrium and the NHSE leads to directed state accumulations that diverge in the thermodynamic limit. However, when the system Hamiltonian is related to its Hermitian adjoint by a similarity transform and thus present a Pseudo-Hermitian symmetry, the partition function is real and we can use traditional thermodynamics in a biorthogonal formalism \cite{gardas2016non}. In this sense, an analysis of the topological phase transitions present in non-Hermitian systems is particularly interesting.

	Although topological phase transitions do not present a local order parameter and cannot be described using a Ginzburg-Landau formalism, a scaling analysis can still be used to characterize them \cite{chen2016scaling, chen2017correlation, chen2018weakly, chen2019universality, molignini2019generating, molignini2020unifying, continentino2017quantum, CONTINENTINO2017A1, griffith2018casimir, rufo2019multicritical, wang2018decoding}. 

	A formalism based on the Hill nanothermodynamics \cite{hill1994thermodynamics}, which is specially suited to analyze thermodynamics of finite size systems \cite{chamberlin99, chamberlin2000mean,chamberlin2015big,latella2015thermodynamics, Li2014, hill2001different, bedeaux2018}, gives a very interesting perspective on topological phase transitions \cite{quelle2016thermodynamic, kempkes2016universalities, van2018thermodynamic, cats2018staircase, arouca2020thermodynamics,yunt2019topological}, despite the intrinsically non-extensive character of topological systems. Due to the bulk-boundary correspondence, the transition at the bulk is related to the one at the surface of the material. As these components have different dimensions, they usually display phase transitions with different orders \cite{quelle2016thermodynamic, kempkes2016universalities, van2018thermodynamic, cats2018staircase, arouca2020thermodynamics}. The nanothermodynamics approach is very robust, being able to describe systems without bulk-boundary correspondence \cite{cats2018staircase}, higher-order topological insulators \cite{arouca2020thermodynamics} and even heat machines \cite{yunt2019topological}. 

	In this work, we study the thermodynamic behavior of a paradigmatic non-Hermitian model, namely the non-Hermitian Su-Schrieffer-Hegger (SSH) model. We calculate the grand potential,  separate it in an extensive (scales with system size) and a non-extensive (does not scales with system size) components and characterize the phase transitions. This is done by analyzing which derivative of the grand potential presents divergences/discontinuities at the critical points,  and by calculating its critical exponents, the electronic density, and the density of states for transitions that occur for finite $\mu$. 

	We find that the transitions between ``Hermitian" phases, \textit{i.e.} phases that can be deformed into a phase of a Hermitian model, belong to the universality class of the Dirac model \cite{chen2017correlation}. On the other hand, transitions between non-Hermitian phases, \textit{i.e.} phases that have no Hermitian counterpart, belong to the non-Hermitian Dirac model universality class and exhibit different critical exponents with respect to the usual Dirac model. Remarkably, the phase transitions in a ``Hermitianized" version of the model belong to the Dirac model universality class, despite the fact that the system then consists of two copies of the non-Hermitian SSH Hamiltonian. 

	The situation for transitions between ``Hermitian" and ``non-Hermitian" phases is completely
 different depending on the boundaries conditions. While for periodic boundary conditions (PBC) they also belong to the non-Hermitian Dirac universality class, for open boundary conditions (OBC) they present a non-Bloch band collapse. This transition is very idiosyncratic because it involves another length scale, besides the correlation length. This new scale, called skin depth, is due to the analytic continuation of the momentum in the complex plane, what makes the bulk modes to be localized. The presence of this length scale, which goes to zero at the phase transition, makes that the correlation function does not spread through the whole material at criticality, but rather is extremely localized in one of the edges! This implies that this phase transition has a very peculiar character: the Josepshon hyperscaling relation is not obeyed and the extensive and non-extensive contributions scale equally. 

	The paper is organized as follows: In Section \ref{sec_thermo}, we review the thermodynamics tools that will be used in this work, with a special focus on scaling ideas in quantum phase transitions. In Section \ref{sec_nh_ssh}, we very briefly present and discuss the non-Hermitian SSH model. Since the phase diagrams are different for PBC and OBC, we separate the results for each case. In Section \ref{sec_PBC}, we discuss the phases for PBC and analyze their thermodynamic behavior and critical exponents. The main finding is that they present critical exponents different from the Hermitian version. Interestingly, if one considers a ``Hermitianized" version of this model, as shown in Section \ref{sec_herm}, one sees that they present the same phase transitions, but with critical exponents equal to the Hermitian SSH model. In Section \ref{sec_OBC}, we consider the system with OBC and its periodic counterpart, the surrogate Hamiltonian. We calculate the thermodynamical behavior, the electronic density and the density of states, and perform an analysis of the critical exponents of the phase transition between ``Hermitian" and between ``non-Hermitian" phases. We finish this Section contextualizing our results in terms of the Hermitian and non-Hermitian Dirac model universality classes. In Section \ref{sec_nBBC}, we focus on the non-Bloch band collapse. As in this system the gap closes into a flat band, we perform a scaling analysis of the correlation length, in addition to the study of the gap closing. This analysis shows that the presence of the skin depth changes completely the scaling relations, and that the Josepshon hyperscaling relation is no longer valid for the non-Bloch band collapse. This corroborates the completely exceptional character of the non-Bloch band collapse as a phase transition.    
	
\section{Thermodynamic Approach to Non-Hermitian Systems}\label{sec_thermo}

	Non-Hermitian systems can possess complex energies and exhibit an extensive sensitivity to boundary conditions. Therefore, it is at first sight questionable whether they can be described thermodynamically. In particular, instabilities due to the non-Hermitian character of the Hamiltonian are amplified at large system sizes, such that systems with different boundary conditions and sizes display very different spectral properties.
		
Regardless of that, the thermodynamics of finite non-Hermitian systems can still be rigorously defined via Hill thermodynamics~\cite{hill1994thermodynamics}, without recourse to the thermodynamic limit. In particular, there exist several classes of non-Hermitian systems with eigenenergies that sum to a real value, either due to the non-Hermitian skin effect, or special symmetries like pseudo-Hermiticity (also called PH in the literature)~\cite{gardas2016non}. 

\subsection{Partition function and grand potential}

A pseudo-Hermitian Hamiltonian $H_{PH}$ is related to its Hermitian adjoint $H_{PH}^\dagger$ by a similarity transformation $U$,	
	\begin{equation}
		H_{PH}^\dagger=UH_{PH}U^{-1},
		\label{eq_PH}
	\end{equation}
which is defined by \cite{gardas2016non}
	\begin{equation}
	\begin{cases}
		U=\sum_{m} \ket{\psi^{L}}\bra{\psi^{L}},\\
		U^{-1}=\sum_{m} \ket{\psi^{R}}\bra{\psi^{R}},
	\end{cases}	
		\label{eq_U}
	\end{equation}
where the $\ket{\psi^{L}}$ and $\ket{\psi^{R}}$ are the left and right eigenvectors\footnote{These relations do not hold at the non-Bloch band collapse point because the Hamiltonian is not diagonalizable. However, they do hold very close to this exceptional point as attested by the scaling analysis performed in Section \ref{sec_nBBC}.}, respectively,
	\begin{equation}
		H_{PH}\ket{\psi^{R}}=E\ket{\psi^{R}}, \qquad H_{PH}^\dagger\ket{\psi^{L}}=E^*\ket{\psi^{L}}. 
		\label{eq_PH_psi}
	\end{equation}
	
	Due to this relation, the energies come in complex conjugated pairs and the  partition function
	\begin{equation}
		Z=\text{Tr}\left[e^{-\beta\left(\hat{H}-\mu \mathcal{N}\right)}\right]
		\label{eq_Z}
	\end{equation}
is real, where $\beta=1/k_B T$, $\mu$ is the chemical potential and $\mathcal{N}$ is the number operator. This real form of $Z$ is amenable to usual thermodynamics approaches, other than the fact that correlation functions should be calculated with the biorthogonal basis $\ket{\psi^{L}}$ and $\ket{\psi^{R}}$, which is equivalent to treating $U$ as a metric operator in the inner product~\cite{gardas2016non}.

	For a free fermionic system with a spectrum $\left\{\epsilon_{m}\right\}$, the grand potential takes the form  
	\begin{equation}
		\Omega\left(T, \mu\right)=-\frac{1}{\beta}\sum\limits_{m}\ln\left[1+e^{-\beta\left(\epsilon_{m}-\mu \right)}\right],
		\label{eq_Omega}
	\end{equation}
which, in the $T=0$ ($\beta\rightarrow \infty$) limit, becomes
	\begin{equation}
		\Omega\left(T=0, \mu\right)=\sum\limits_{\Re\epsilon_m\leq \mu}\left(\epsilon_{m}-\mu\right).
		\label{eq_Omega_T=0}
	\end{equation} 
Notice that for non-Hermitian systems, the filling of the Fermi sea is determined by the real part of the spectrum~\cite{herviou2019entanglement}.

	For $1D$ systems, the grand potential in the thermodynamic limit usually contains an extensive part $\omega_\text{ext}L$, where $L$ is the size of the system, and a non-extensive part $\omega_\text{n-ext}$, which are related to the bulk and edge energy contributions, respectively \cite{quelle2016thermodynamic, kempkes2016universalities} . As a result, one can decompose $\Omega$ as
	\begin{equation}
		\Omega=\omega_{\text{ext}}L+\omega_{\text{n-ext}}.
		\label{eq_Omega_L}
	\end{equation}
	
	This scaling, however, does not hold in general. In the presence of non-local correlations, other type of scalings can appear in the grand potential \cite{cats2018staircase}. For non-Hermitian systems, the spectrum for OBC can be obtained by a non-local mapping of the Bloch Hamiltonian \cite{lee2019unraveling, kawabata2019symmetry}, as explained in Subsection \ref{subsec_surr}. Therefore, it was not a priori clear whether this scaling with $L$ should hold for this system \cite{li2020critica, li2020impurity}. Nevertheless, this turned out to be the case for all the phases of the model that we considered in this work, as shown in Appendix \ref{app_scaling_L}. 

	For most topological systems, a bulk-boundary correspondence is present and one can obtain these two contributions by calculating the grand potential with different boundary conditions~\cite{cats2018staircase, arouca2020thermodynamics}. The system with PBC only has the bulk contribution, while the system with OBC presents both contributions. We can then obtain the boundary contribution by subtracting the grand potential calculated for PBC from that with OBC. 
	
	 This approach, however, should be modified when the non-Hermitian skin effect invalidates the bulk-boundary correspondence. In this case, one should explicitly vary the system size $L$~\cite{quelle2016thermodynamic, kempkes2016universalities, van2018thermodynamic}, evaluate the grand potential for many values of $L$, and do a linear fit to separate the extensive and non-extensive contributions to $\Omega$. 

%	\subsection{Thermodynamic responses using correlation functions}\label{subsec_corr}

	\subsection{Scaling near a quantum phase transition}\label{subsec_scaling}

	The main focus of this work is to show, via a thermodynamic approach, unconventional scaling properties around non-Hermitian critical points, especially when boundary effects scale anomalously. To do that, we review the framework of scaling functions and critical exponents~\cite{continentino2017quantum, *continentino1994quantum}.

	If we consider a reduced parameter $g$ that tunes a critical phase transition, such that it occurs at $g=0$, we can associate critical exponents~\cite{continentino2017quantum, sachdev_2011, vojta2003quantum} to different thermodynamic quantities. The grand potential itself presents such scaling. Let us decompose (any component of) the reduced grand potential $\omega_{j}$, $j=\text{ext}/\text{n-ext}$ as a sum of a regular part $\omega_{\text{reg}}$ and a singular $\omega_{s}$ part 
	\begin{equation}
		\omega_j\left(g\right)=\omega_{reg}\left(g\right)+\omega_s\left(g\right),
		\label{eq_omega_s+reg}
	\end{equation}
such that the singular part scales with $g$ close to the phase transition
	\begin{equation}
		\omega_{s}\left(g\right)\sim \left|g\right|^{2-\alpha},
		\label{omegasg}
	\end{equation}
where $\alpha$ is the canonical critical exponent. $2-\alpha$ is then the order of the phase transition because derivatives of $\omega_{s}$ of orders higher than $2-\alpha$ are discontinuous at the phase transition point.

We consider a quantum setting where occupied states with momenta $p$ \footnote{We adopt $p$ instead of $\mathbf{p}$ to be consistent with the notation used in our results, as we are considering a $1D$ chain.} are separated by a gap $\Delta$, according to the power-law scaling relations
	\begin{eqnarray}
		\Delta\left(p, g=0\right)\propto \left|p\right|^z\label{eq_z},\\
		\Delta\left(p=0, g\right)\propto \left|g\right|^{\nu z}\label{eq_nu},
	\end{eqnarray}
where $p$ is the momentum measured with respect to the point $\mathbf{K_c}$ at which the gap closes. The associated critical exponents $z$ and $\nu$ corresponding to the scaling of the gap with the momentum $p$ and with $g$, respectively, are related to the order of the phase transition as follows.	For each real energy interval $d\epsilon$, there are $|\bold p|^d$ occupied states. Hence $\omega_s=\int^\Delta|\bold p|^d d\epsilon =\int^\Delta \epsilon^{d/z}d\epsilon \propto \Delta^{1+d/z}\propto |g|^{\nu(d+z)}$. Comparing with Eq.~\ref{omegasg}, we obtain
		\begin{equation}
		2-\alpha=\nu\left(d+z\right),
		\label{hyperscale}
	\end{equation}
the Josephson's hyperscaling relation \cite{continentino1994quantum}, which can be derived from purely dimensional arguments on a critical system. Hence, by studying how the gap closes, one can determine $z$ and $\nu$, and establish the order of the phase transition for a system of dimensionality $d$.

	This relation also implies that the same system in settings with different dimensionalities undergo different orders of phase transition, even when their critical exponents $\nu$ and $z$ are the same. For the Hermitian SSH model for instance, $z=1$ and $\nu=1$, which yields a first-order phase transition for its edge (non-extensive contribution, $d=0$) and a second-order transition for its bulk (extensive contribution, $d=1$)~\cite{kempkes2016universalities}. Below, we shall see that the non-Hermitian variant of the SSH model possess much more intricate bulk-vs-boundary scaling behavior.

	\subsection{Thermodynamics of topological systems}\label{subsec_top}

		Topological phase transitions are transitions between inequivalent phases, distinguished by different topological invariants. The change of topology is usually accompanied by a gap closing. In general, a gap closing signals a quantum phase transition due to the change of the system ground state, thus leading to the scaling behavior discussed in Subsection \ref{subsec_scaling}. In this sense, a natural question that arises is whether these two notions of phase transitions are equivalent. 

	Many works indicate that topological phase transitions show distinct scaling behavior ~\cite{CONTINENTINO2017A1, griffith2018casimir, rufo2019multicritical, molignini2019generating, chen2019universality, chen2018weakly, chen2016scaling, molignini2020unifying}. In addition, the thermodynamic approach used here ~\cite{quelle2016thermodynamic, kempkes2016universalities, van2018thermodynamic, cats2018staircase} has revealed that when the change of a topological invariant occurs concomitantly with a drastic variation of the spectrum, the grand potential (or its derivatives) exhibit a discontinuity at the phase transition. Then, the two notions of phase transitions are equivalent. 

	There are, however, some topological phase transitions that do not show signatures in the grand potential. An example are boundary-obstructed topological phases \cite{benalcazar17_PRB, benalcazar2017_Science,khalaf2019boundary}, for which the change of topology is revealed by the Wannier spectrum, and no signature is visible in the grand potential~\cite{arouca2020thermodynamics}. 

%	Moreover, for finite temperature, the There are some examples where the thermodynamic grand potential is clearly defined and can be used to define the phase diagram of the system and other  approach have some advantages. The grand potential is clearly defined at both zero and finite temperature, while there are some subtleties in defining topological invariants for finite temperature~\cite{viyuela14_uhlmann, budich15_top_dens}. In a previous work, it was shown that the definitions of invariants for zero and finite temperature give the same phase diagram that the grand potential~\cite{kempkes2016universalities}. Furthermore, topological effects can play an important whole in thermodynamic processes~\cite{yunt2019topological, ott20_radiative}.  

\section{Non-Hermitian SSH Model}\label{sec_nh_ssh}

\begin{figure}[!htb]
		\centering
		\includegraphics[width=\linewidth]{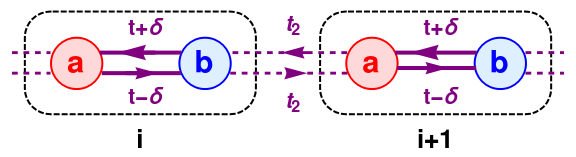}
		\caption{Sketch of the model of Eq.~\ref{eq_ham}. The intracell (solid lines) hopping parameters are non-reciprocal (one is $t+\delta$, while the other is $t-\delta$), which leads to the non-Hermitian character of this model. The intercell hopping (dashed lines) is $t_2$. The ratios $t/t_2$ and $\delta/t_2$ determine the phases of the system.}
		\label{fig_lattice}
	\end{figure}

	One of the simplest models exhibiting both topological effects and the NHSE, which breaks the bulk-boundary correspondence, is the non-Hermitian variant of the SSH model~\cite{kunst2018biorthogonal, yao2018edge,yin2018geometrical, lee2019anatomy, kawabata2019symmetry, ghatak2019observation, flebus2020non, helbig2020generalized}. There are basically two ways to make the SSH model non-Hermitian. One is to introduce a complex intracell hopping, which is equivalent to introducing a Peierls phase in the intracell bonds. As this phase can be gauged away, this system still obeys bulk-boundary correspondence, despite exhibiting complex energies. Another approach, which will be the one under investigation in this work, is to introduce a real non-reciprocal intracell hopping. This leads to the pseudo-Hermitian Hamiltonian
	\begin{align}
	\begin{split}
		H=&\left(t+\delta\right)\sum\limits_{i}a^\dagger_{i}b_{i}+\left(t-\delta\right)\sum\limits_{i} b^\dagger_{i}a_{i}\\
		&+t_2 \sum\limits_{i} \left(a^\dagger_{i+1} b_{i}+h.c.\right),
		\label{eq_ham}
	\end{split}
	\end{align}
where the non-reciprocal intracell hopping parameters are given by $t-\delta$ and $t+\delta$, and the intercell hopping parameter is denoted by $t_2$ . A sketch of this model is shown in Fig.~\ref{fig_lattice}. A nonzero value of $\delta$ breaks the Hermiticity of this Hamiltonian and leads to the NHSE i.e. a dramatic modification of all eigenstates upon a local change in the value of the hoppings at the boundary \cite{lee2019anatomy}.

	The ratios $t/t_2$ and $\delta/t_2$ determine the different topological phases of this system, which include phases that are not possible in Hermitian systems. These phases become completely different upon varying the boundary conditions, as the bulk-boundary correspondence is not respected. Hence, we will consider the different boundary conditions separately. As this system is pseudo-Hermitian, with complex conjugate pairs of eigenenergies, the traditional thermodynamic approach applies and we can characterize its phase transitions using the grand potential.
	
\section{Periodic Boundary Conditions}\label{sec_PBC}

\subsection{Phase Diagram}\label{subsec_phase}

	We start our analysis by reviewing the non-Hermitian SSH system with PBCs. The Bloch Hamiltonian corresponding to Eq.~\eqref{eq_ham} \cite{kawabata2019symmetry} is given by
	\begin{equation}
		h\left(k\right)=\left(\begin{matrix}0&t-\delta+t_2e^{-i  k}\\t+\delta+t_2e^{+i  k}&0\end{matrix}\right).
		\label{eq_bloch_ham_nh_ssh}
	\end{equation}
	\begin{figure}[!htb]
		\centering
		\includegraphics[width=\linewidth]{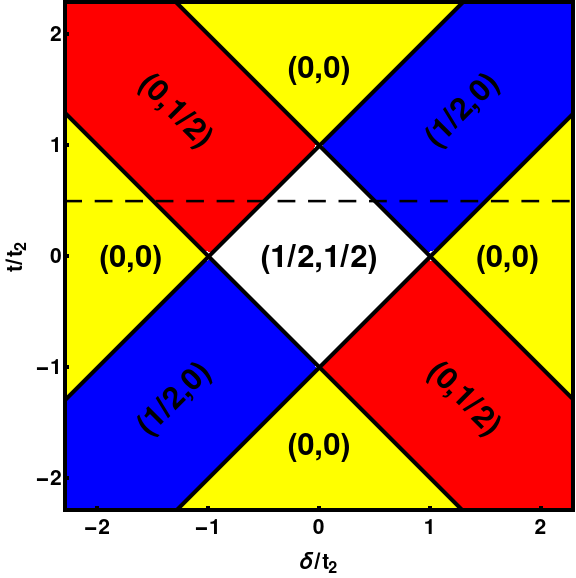}
		\caption{Phase diagram for the non-Hermitian SSH system with periodic boundary conditions. The phases are characterized by the winding numbers $(W_1, W_2)$. Hermiticity requires $W_1=W_2$, so the Hermitian phases are given by $(0, 0)$ and $(1/2, 1/2)$. On the other hand, the non-Hermitian phases are $(0,1/2)$ and $(1/2, 0)$. In our study, we consider a fixed value of $t=0.5$ and vary $\delta$ (dashed line) to explore all the phases transitions. This phase diagram was originally obtained in Ref.~\cite{yin2018geometrical}.}
		\label{fig_phase_diagram_pbc}
	\end{figure}
	By diagonalizing this Hamiltonian, we find the energy levels for PBC:
	\begin{equation}
	\begin{split}
		\epsilon_\pm \left(k\right)&=\pm\epsilon \left(k\right)\\
			&=\pm \sqrt{1^2+t^2-\delta^2+2 t \cos k-2i  \delta \sin k},
		\label{eq_spectrum}
	\end{split}
	\end{equation}
where we have set the characteristic energy scale $t_2=1$ and the lattice parameter length scale $a=1$ (this make that $k$ goes from $-\pi$ to $\pi$).

 	The system exhibits a gap closing when
	\begin{equation}
		1\pm 2 t+t^2-\delta^2=0\rightarrow \delta_c=\pm \left|1\pm t\right|.
		\label{eq_delta_c_PBC}
	\end{equation}
	This occurs at $k=0$ for $\delta_c=1+t$ and $k=\pi$ for $\delta_c=1-t$, which separates the system into different topological phases.
	
	To understand the nature of these phases, we recast the Bloch Hamiltonian \eqref{eq_bloch_ham_nh_ssh} as a $\bold d=(d_x,d_y)$ vector multiplied by the Pauli matrix basis $\sigma=\left(\sigma_x,\sigma_y\right)$,
	\begin{equation}
		h\left(k\right)=d_x\left(k\right)\sigma_x+d_y\left(k\right)\sigma_y,
		\label{eq_Dirac}
	\end{equation}
with $d_x$ and $d_y$ given by
	\begin{equation}
	\begin{cases}
		d_x\left(k\right)=t+t_2\cos\left(k\right)\\
		d_y\left(k\right)=t_2\sin\left(k\right)-i  \delta.
	\end{cases}
	\end{equation}
In this case, the topological phases may be identified by the associated winding numbers~\cite{yin2018geometrical}
	\begin{equation}
		\mathbf{W}=(W_1, W_2),
	\end{equation}
where $W_{i}$, $i=1,2$ are defined~\cite{yin2018geometrical} as
	\begin{equation}
		W_i=\frac{1}{2\pi}\oint \partial_k \arctan\left[\frac{\Re d_y\left(k\right) \pm \Im d_x\left(k\right) }{\Re d_x\left(k\right) \mp \Im d_y\left(k\right) }\right] dk,
	\end{equation}	
with the upper sign for $i=1$ and the lower sign for $i=2$.

	As $k$ varies, the trajectory of $\bold d$ in the $\Re{d_x}$/$\Re{d_y}$ plane can be topologically nontrivial if it encircles exceptional points (EP) at $\left(\pm \Im{d_y},\mp \Im{d_x} \right)$, where $W_{i}$ is ill defined. Hence, $W_{i}$ is the winding number around the EPs, which characterizes the phases of this system, and Eq.~\eqref{eq_delta_c_PBC} provides the values for $\delta$ that mark the transition between states with different $W_{i}$, thus leading to the phase diagram given in Fig.~\ref{fig_phase_diagram_pbc}. For Hermitian systems ($\delta=0$), both components of this winding number should be equal, so $\mathbf{W}=(0,0)$ is equivalent to the Hermitian trivial phase and $\mathbf{W}=(1/2, 1/2)$ to the Hermitian topological phase. In addition, we have the exclusively non-Hermitian topological phases $(0,1/2)$ and $(1/2,0)$.

	\subsection{Thermodynamics}\label{subsec_thermo_pb}	
	\begin{figure}[!htb]
%		\hspace*{-0.5cm}     
		\includegraphics[width=1.0\linewidth]{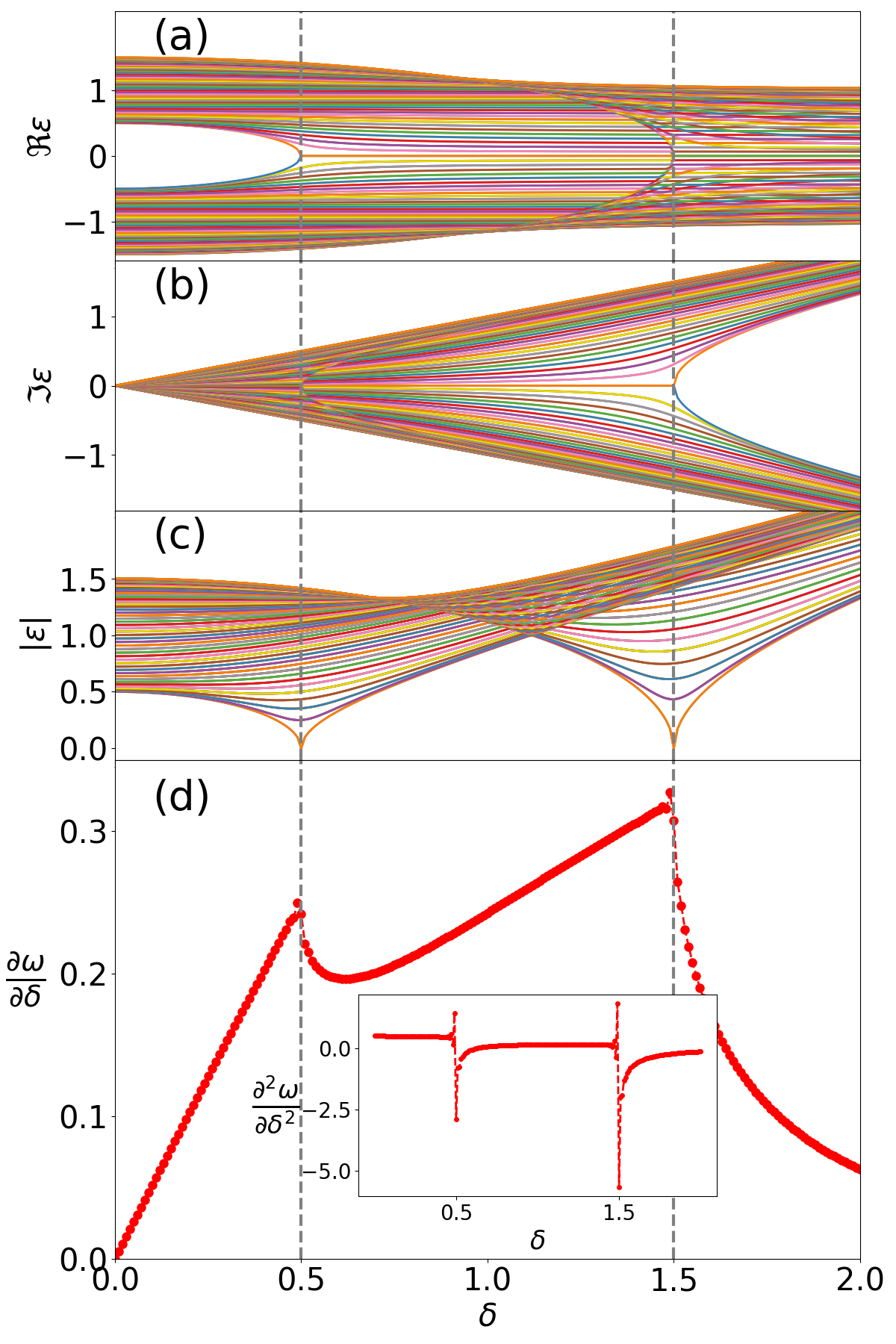}
		\caption{Spectrum and derivatives of the grand potential of the non-Hermitian SSH model as a function of $\delta$ for PBC, $t=0.5$, and $\mu=0$. (a) real, (b) imaginary and (c) absolute value components of the energy. (d) First derivative of the grand potential. Inset: second derivative of the grand potential, which is discontinuous at the two critical points. We use $100$ discretization values of $k$ for the energy simulation and $1000$ values of $k$ for the grand potential. The gray vertical dashed lines indicate the (theoretical) critical values of $\delta$.}
		\label{fig_pbc}
	\end{figure}

Since the eigenenergies of the system determine the free energy, phases with different real, imaginary and absolute values of the energy spectrum will give rise to different thermodynamic properties, particularly across phase transitions. Below, we detail the spectral behavior of the system as $\delta$ is varied, to concretely explain how various critical scenarios manifest as kinks in the grand potential. 

The results for $t=0.5$ are shown in Fig.~\ref{fig_pbc} (a)-(c). This value is chosen to span all phases upon varying $\delta$. The spectrum shows a gap closing (merging of bands) at $\delta_c=0.5$ and a merging of bands (gap closing) at $\delta_c=1.5$ for the real (imaginary) component of the energy. The absolute value of the energy shows a very distinct gap closing at both values of $\delta_c$. %In this way, the spectrum can be used as an indicator of these phase transitions.

	The signatures of the phase transitions seen in the energy spectrum are also manifest in the derivative of the reduced grand potential $\omega$ \footnote{Notice that as the system with PBC does not have an edge, the grand potential is extensive, such that $\omega=\Omega/L$ and we do not need to define extensive and non-extensive contributions. This will also be the case for the surrogate Hamiltonian, as it is an analytical continuation of the PBC Hamiltonian.}, which indeed shows kinks at $\delta_c=0.5$ and $\delta_c=1.5$ (see Fig.~\ref{fig_pbc}), with its second derivative showing discontinuities. One could identify the order of the phase transition by applying the Ehrenfest classification~\cite{jaeger}, which associates the order of the phase transition to the order of the derivative of the grand potential that diverges or exhibits a discontinuity. In this case, one would conclude that both phases transitions have the same order. However, the precise order of these phase transitions do not seem to be unequivocally determined by only considering the discontinuities at the critical point. To acquire a clearer picture, we will calculate the critical exponents involved in the Josepshon's hyperscaling relation [Eq.~\eqref{hyperscale}].

	\subsubsection{Critical Exponents}

	\begin{figure}[!htb]
		\centering
		\includegraphics[width=\linewidth]{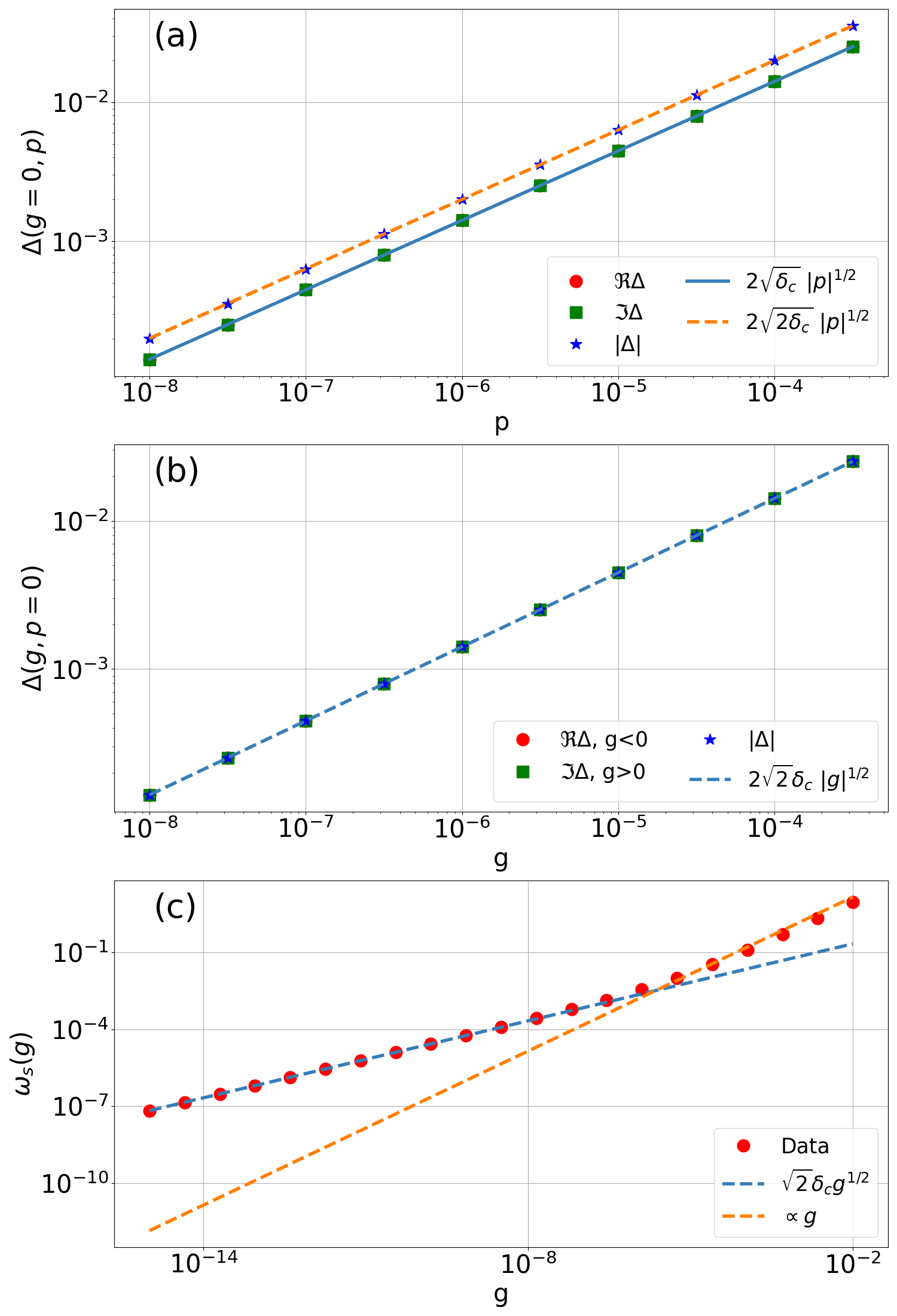}
		\caption{Scaling analysis for the gap $\Delta$ closing and for the singular part of the grand potential $\omega_s$ near the critical point $\delta_c=0.5$ and $K_c=\pi$ for $t=0.5$. (a) In momentum space, the real (red circles) and imaginary (green squares) parts of the gaps are very well described by Eq.~\eqref{eq_Delta_p} (blue solid line), while its absolute value (blue stars) is described by the first of Eqs.~\eqref{eq_delta_abs} (orange dashed line). This result confirms the critical exponent $z=1/2$.(b) For the closing of the gap in the parameter space, the real (red circles) and imaginary (green squares) components are very well described by Eq.~\eqref{eq_Delta_g} (blue solid lines) and the absolute value (blue stars) is determined by Eq.~\eqref{eq_delta_abs}(orange dashed lines). This result confirms the critical exponent $\nu=1$. (c) The scaling of $\omega_s$ [red circles, obtained using Eq.~\eqref{eq_omega_sing}] goes with $g^{1/2}$, which confirms that $2-\alpha=1.5$ (see discussion in the text). For values of $g$ further away from the phase transition, there is a subdominant scaling with $g^{1}$, similar to the one in the Hermitian SSH model. We used $\mu=0$, $50$ discretization $k$ values for $\Delta$, and $1000$ values of $k$ for $\omega_s$.}
		\label{fig_scaling_pb}
	\end{figure}  

	An examination of how the gap closes in both the momentum and parameter space allows us to obtain the critical exponents $\nu$ and $z$ defined in Eqs.~\eqref{eq_z} and \eqref{eq_nu}. As this system has only two bands, the gap can be obtained from Eq.~\eqref{eq_spectrum}
		\begin{equation}
			\Delta\left(k\right)=\epsilon_+\left(k\right)-\epsilon_-\left(k\right)=2\epsilon\left(k\right).
		\end{equation}
	
	If we write the spectrum in Eq.~\eqref{eq_spectrum} in terms of the reduced parameter 
	\begin{equation}
		g=\frac{\delta-\delta_c}{\delta_c}\rightarrow \delta=\left(1+g\right)\delta_c
	\end{equation}
and the relative momentum $p=k-K_c$, we find that near the critical point ($g\ll 1$ and $p\ll 1$), the gap closes as
	\begin{equation}
		\Delta\left(g,p\right)=2\sqrt{2}\sqrt{-\delta_c^2 g+ i  \delta_c p}.
		\label{eq_Delta_crit}
	\end{equation}
	If we calculate this expression now at $g=0$ and $p=0$, we find the critical exponents $z$ and $\nu$, respectively. For the $z$ exponent,
	\begin{align}
	\begin{split}
		\Delta\left(g=0,p\right)=2\sqrt{2\delta_c}\sqrt{i  p}= 2\sqrt{\delta_c}\left(1+i \right) \left|p\right|^{1/2}.
		\label{eq_Delta_p}
	\end{split}
	\end{align}

	We observe that both the real and imaginary parts have the same critical exponent, $z=1/2$, and the same critical amplitude, $2\sqrt{\delta_c}$, on both sides of the phase transition.

	On the other hand, looking at how the gap closes in terms of $g$ for $p=0$, we find			
	\begin{align}
	\begin{split}
		\Delta\left(g,p=0\right)=2\sqrt{2}\delta_c\sqrt{-g}=2\sqrt{2}\delta_c \left|g\right|^{1/2}\sqrt{-\frac{g}{\left|g\right|}}.
		\label{eq_Delta_g}
	\end{split}
	\end{align}
	As $\sqrt{-g/\left|g\right|}=1$ for $g<0$ and $\sqrt{-g/\left|g\right|}=i $ for $g>0$, the critical exponent becomes ill defined for positive (negative) values of $g$ for the real (imaginary) part of the gap as the critical amplitude is equal to zero.

	If instead we consider how the absolute value of the gap behaves,	
	\begin{align}	
	\begin{split}
		\left|\Delta\left(g,p\right)\right|=&2\sqrt{2}\sqrt[4]{\delta_c^4 g^2+\delta_c^2 p^2},
	\end{split}
	\end{align}
we obtain	
	\begin{align}	
	\begin{cases}
		\left|\Delta\left(g=0,p\right)\right|=2\sqrt{2\delta_c}\left|p\right|^{1/2}\\
		\left|\Delta\left(g,p=0\right)\right|=2\sqrt{2}\delta_c\left|g\right|^{1/2},
	\end{cases}
		\label{eq_delta_abs}
	\end{align}
and $\nu_{\pm}=1$. In this way, we understand that this apparent discrepancy in the critical exponents for one side or the other of the phase transition is only because of considering the real and imaginary parts of the gap separately. For the absolute value, the behavior is symmetric around the critical point. 

 	We can check that this is indeed how the gap closes by numerically investigating the particular critical point $\delta_c=0.5$ and $K_c=\pi$ for $t=0.5$ in Fig.~\ref{fig_scaling_pb} (a)-(b), which confirms that close to the phase transition (both in momentum and parameter space), the gap has a power-law behavior, with the critical exponents and amplitudes given by Eqs.~\ref{eq_Delta_p}, \ref{eq_Delta_g} and \ref{eq_delta_abs}.

	If we use these critical exponents in the Josepshon hyperscaling relation, we obtain 
	\begin{equation}
		2-\alpha=1\left(1+\frac{1}{2} \right)=1.5.
	\end{equation}
	This implies that the non-Hermitian SSH model has a critical phase transition of fractional order, and belongs to a different universality class as compared to the Hermitian SSH model \cite{chen2017correlation}.
	
	 To confirm this behavior, we perform a scaling of the grand potential with $g$ close to the point of the phase transition. First, one needs to extract the singular part of the grand potential. This contribution is usually \cite{cardy1996scaling} obtained by
	\begin{equation}
		\omega_{s}\left(g\right)=\omega\left(g\right)-\omega\left(0\right),
		\label{eq_omega_sing}
	\end{equation}
as $\omega_{\text{reg}}\left(g\right)\approx \omega_{\text{reg}}\left(0\right)$ and $\omega_s\left(0\right)=0$ ($2-\alpha>0$ in general).

	If one considers the real or imaginary part of the spectrum, instead of the its absolute value, the gap closing occurs only for a given sign of $g$ . As the free energy is calculated by the real values of $\epsilon$, this will lead to different critical exponents for different sides of the phase transition. To remediate this problem, we calculate the singular part of the grand potential by
	\begin{equation} 
		\omega_{s}\left(g\right)=\omega\left(g\right)-\omega\left(-g\right),
	\end{equation}
where we assumed that $\omega_{\text{reg}}\left(g\right)\approx \omega_{\text{reg}}\left(-g\right)$ for small enough $g$ and that $\omega\left(g\right)\approx \omega_{\text{reg}}\left(g\right)$ for one of the sides of the transition.

	\begin{figure}[!htb]
		\centering
		\includegraphics[width=\linewidth]{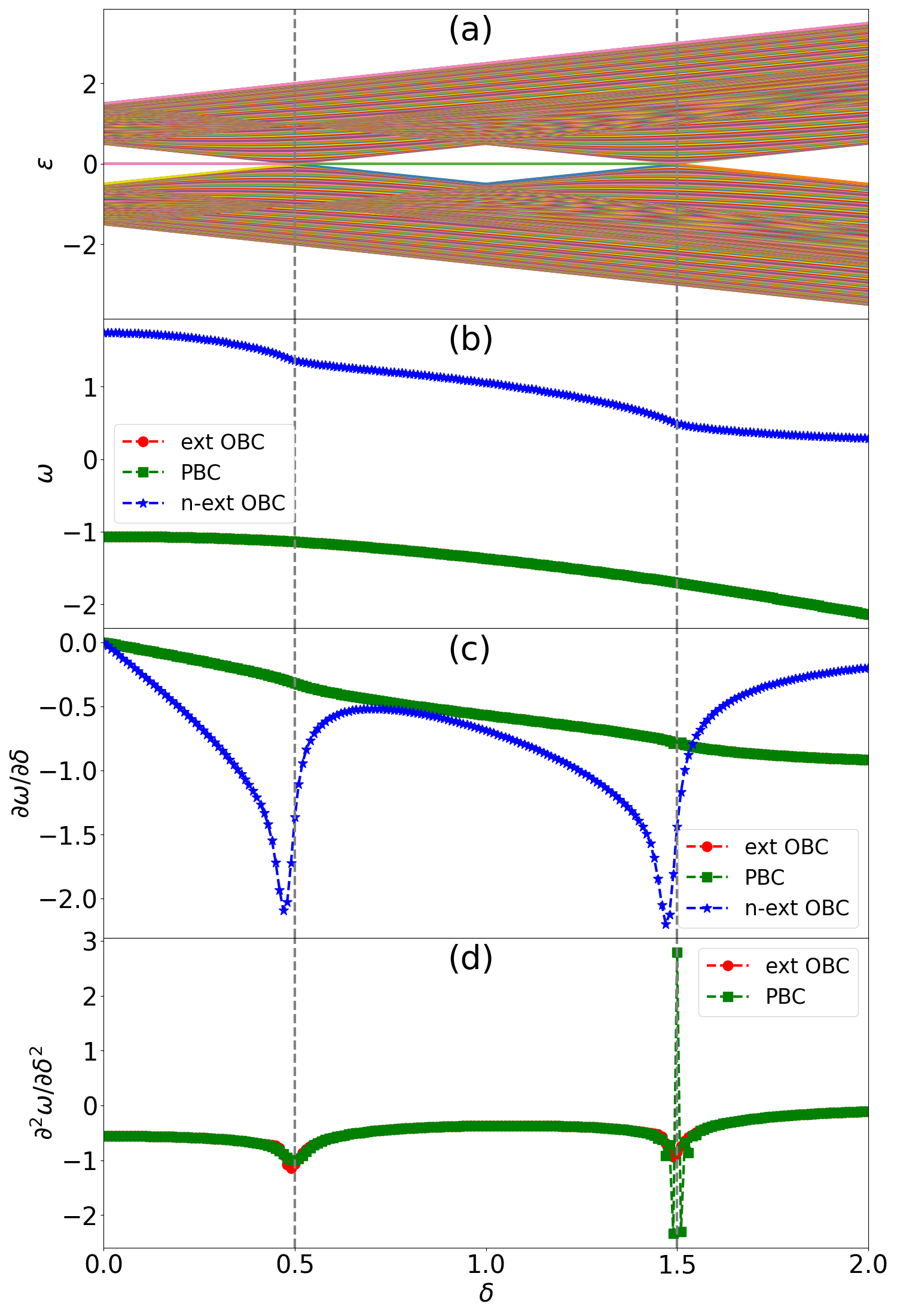}
		\caption{Spectrum and thermodynamic properties of $\mathcal{H}$ for $t=0.5$. (a) The spectrum shows a number of zero modes equal to $W_1+W_2$, which are a signature of the phase transitions in the phase diagram in Fig.~\ref{fig_phase_diagram_pbc}. These transitions are also shown in (b) the grand potential and (c) its first and (d) second derivative. These quantities were calculated for the extensive (red circles with red dashed lines) and the non-extensive (blue stars with blue dashed lines) contributions, as well for PBC (green squares with green dashed lines). We used a system with $100$ unit cells for OBC and with $100$ discretized values of $k$ for PBC  in these simulations. The gray vertical dashed lines indicate the (theoretical) critical values of $\delta$.}
		\label{fig_herm}
	\end{figure}
	Furthermore, to numerically analyze the critical behavior of the grand potential, we need to scale every length by $\xi$ (or, equivalently, $g^{-\nu}$) and every energy scale by the gap (or, equivalently $\xi^{-1/z}$ or $g^{\nu/z}$). Therefore, $\omega_s$ should be multiplied by a factor $\xi$, as it is divided by the system size $L$. Hence, the scaling of this quantity is actually
	\begin{align}
	\begin{split}
		\omega_s\left(g\right) \propto & g^{2-\alpha}\xi\\
		=&g^{2-\alpha}g^{-\nu}=g^{2-\alpha-\nu}\\
		=&g^{1/2}. 
	\end{split}
	\end{align}
	The results for the simulation of this quantity are given in Fig.~\ref{fig_scaling_pb} (c) for $\delta_c=0.5$ and $L=1000$. It is interesting to observe that very close to the phase transition, they follow the expected behavior, but for higher values of $g$ they start to scale with $g^1$, which is the behavior of the Hermitian SSH model. For larger values of $L$, the onset of the scaling with $g^{1}$ happens at smaller $g$ because the scaling actually goes like $\xi g$ and $\xi$ increases with system size. Interestingly, the fact that $z=1/2$ makes  $2-\alpha=1.5$ and justifies why it was difficult to resolve whether the transition was of first or second order. 

	\section{``Hermitianized" SSH model}\label{sec_herm}
	
To connect between the Hermitian and non-Hermitian SSH models, and, in particular, to trace the origin of the fractional order phase transition, we extend the non-Hermitian SSH model to an extended Hermitian Hamiltonian $\mathcal{H}$ comprising $H$ and its conjugate,
	\begin{equation}
		\mathcal{H}=
			\left(\begin{matrix}
			0&H\\
			H^\dagger&0\\	
			\end{matrix}\right).
		\label{eq_Ham_herm}
	\end{equation}
	It was shown in Refs.~\cite{yin2018geometrical, gong2018topological} that this mapping basically transforms each point gap present in the spectrum of the non-Hermitian system into a line gap in the spectrum of $\mathcal{H}$. In this way, the usual bulk-boundary correspondence is recovered in this ``Hermitianized" system. In addition, the number of edge modes in $\mathcal{H}$ is equal to $W_1+W_2$ calculated the Bloch Hamiltonian on Eq.~\eqref{eq_bloch_ham_nh_ssh}  \cite{yin2018geometrical, gong2018topological}.
	
	\begin{figure}[!htb]
		\centering
		\includegraphics[width=\linewidth]{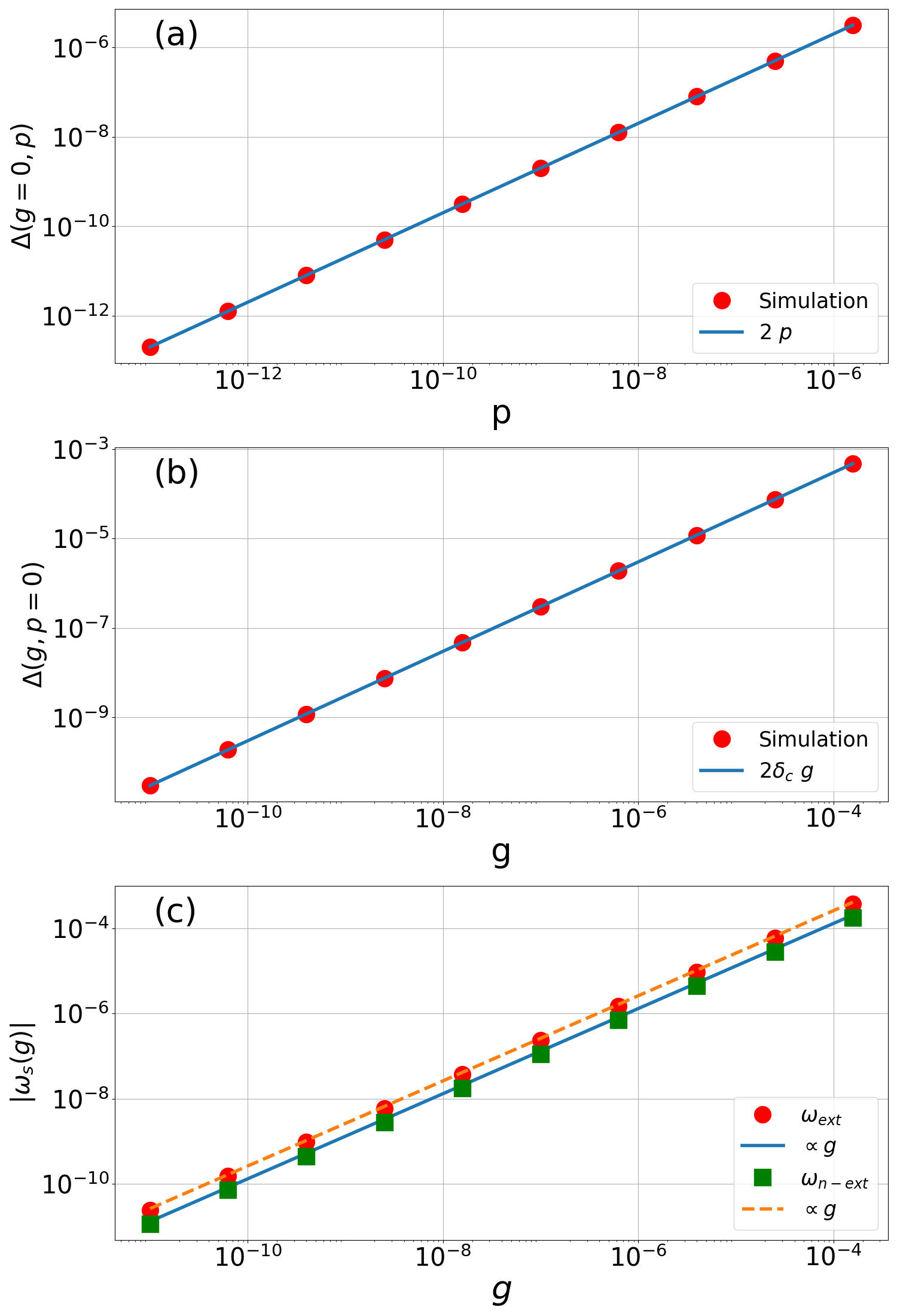}
		\caption{Critical behavior of $\mathcal{H}$ close to the critical point $\delta_c=1.5$, $K_c=0$. The gap closing in both, (a) the momentum and (b) the parameter space exhibits critical exponents $z=1$ and $\nu=1$, respectively, which are the same for the Hermitian SSH model. (c) The singular contribution of the grand potential $\omega_s$ scales with $g$ (solid blue and orange dashed lines. The two different lines represent different amplitudes, both for $\omega_\text{ext}$ (red circles) and $\omega_\text{n-ext}$ (green squares). We used $100$ unit cells for this simulation. The results for the $\delta_c=0.5$ critical point are similar, with the difference that there the extensive and non-extensive contributions are precisely equal.}
		\label{fig_crit.5_herm}
	\end{figure}  
	
	Fig.~\ref{fig_herm} shows the thermodynamic results, together with the spectrum for $\mathcal{H}$, obtained using Hamiltonian \eqref{eq_ham} for $t=0.5$. The spectrum, depicted in Fig.~\ref{fig_herm} (a), shows signatures for both values of $\delta_c$. Upon increasing $\delta$, at $\delta_c=0.5$, one of the zero modes (pink) goes into the bulk and at $\delta_c=1.5$ the other zero mode (green) goes into the bulk. These phase transitions can be clearly seen in the grand potential and its derivatives, as shown in Figs.~\ref{fig_herm} (b)-(d). If we separate the grand potential using Eq.~\eqref{eq_Omega_L}, we find that the extensive contribution $\omega_\text{ext}$ is equal to the one calculated using PBC. This is a distinctive feature of this system, as it recovers the bulk-boundary correspondence. We notice also that the phase transition for the non-extensive contribution $\omega_\text{n-ext}$ is of first order, while the one for $\omega_\text{ext}$ is of second order \footnote{The apparent divergence in the PBC component is due to numerical instabilities diagonalizing the system in the vicinity of the critical point for PBC.}, as in the Hermitian SSH model \cite{kempkes2016universalities}! Despite still containing the non-reciprocal hoppings, this Hermitianized model no longer exhibits fractional order phase transitions. We turn now to an analysis of the critical exponents to verify the order of the phase transitions. 
	
	The gap closes linearly both in momentum and parameter space, as shown in Figs.~\ref{fig_crit.5_herm} (a) and (b), so that $\nu=1$ and $z=1$, like in the Hermitian SSH model \cite{kempkes2016universalities, van2018thermodynamic}. The Josephson hyperscaling relation then becomes $2-\alpha=d+1$, so that it has the same critical behavior of a Hermitian SSH, which explains the previous results.

	To check the scaling of the grand potential close to the critical point, we extract its singular contribution using Eq.~\eqref{eq_omega_sing}, but separating the extensive and non-extensive contributions using Eq.~\eqref{eq_Omega_L}. The scaling of the two contributions $\omega_\text{ext}$ and $\omega_\text{n-ext}$ are given by 
	\begin{equation}
	\begin{cases}
		\omega_\text{ext}^{s}\left(g\right)\propto g^{2}\xi^{1}=g^{2}g^{-1}=g\\
		\omega_\text{n-ext}^{s}\left(g\right)\propto g^{1}\xi^{0}=g,
	\end{cases}
	\end{equation}
as the extensive part scales with $L^{1}$, while the non-extensive one scales with $L^{0}$. This kind of behavior is indeed observed in the simulations, as shown in Fig.~\ref{fig_crit.5_herm} (c).

%\newpage

	\section{Open Boundary Conditions}\label{sec_OBC}
		The system with OBC has very different properties, as compared to the one with PBC. In particular, in the OBC case, there are no bulk modes only skin modes localized at the edges of the system, which implies that the bulk-boundary correspondence is broken via the NHSE.
	\subsection{Phase Diagram}\label{subsec_phase_ob}

	\begin{figure}[!htb]
		\centering
		\includegraphics[width=\linewidth]{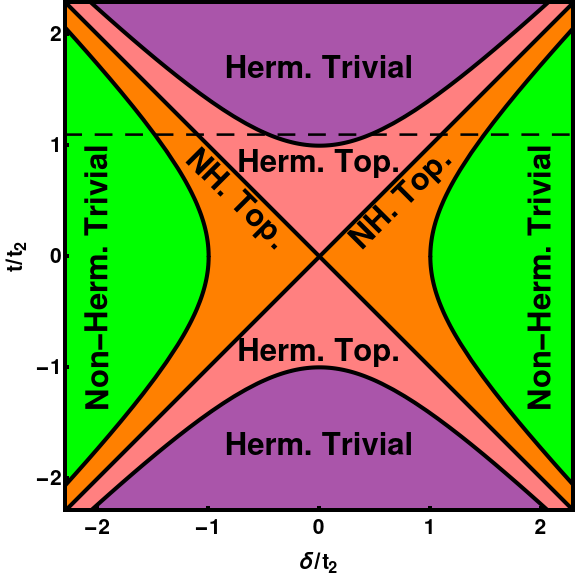}
		\caption{Phase diagram of the non-Hermitian SSH model with open boundary conditions. For $\left|\delta\right|<t$, the phases are continuously connected to the phases of the Hermitian ($\delta =0$) SSH model, and are denoted as Hermitian trivial (``Herm. Trivial", purple) and Hermitian topological (``Herm. Top.", pink) even though they are still non-Hermitian. The boundary between these phases is given by the curves $\left|\delta_c\right|=\sqrt{t^2-t_2^2}$. For $\left|\delta\right|>t$, the phases cannot be deformed to the ones in the Hermitian model, so they are called non-Hermitian trivial (``Non-Herm. Trivial", green) and non-Hermitian topological (``NH. Top.", orange). The boundary between these phases is determined by the curves $\left|\delta_c\right|= \sqrt{t^2+t_2^2}$. For $\left|\delta_c\right|=t$, the non-Bloch band collapse critical line separates the Hermitian and non-Hermitian phases. In our studies, we analyze the phase transitions along the path $t=1.1$ (black dashed line), such that by varying $\delta$, we span all phases.}% This phase diagram was originally obtained in Ref.~\cite{kawabata2019symmetry}.}
		\label{fig_phase_diagram_obc}
	\end{figure}
	
	The phase diagram of the same non-Hermitian SSH model, but with OBC is also very different, see Fig.~\ref{fig_phase_diagram_obc}. For $\left|\delta\right|<t$, the system is adiabatically connected to the Hermitian model and, as such, we denote these phases as  Hermitian trivial and Hermitian topological, depending on their winding number \cite{kawabata2019symmetry}
	\begin{equation}
		W=\oint \frac{dk}{4\pi i}\Tr\left[\sigma_z h^{-1}\left(k\right)\frac{d h\left(k\right)}{dk}\right],
		\label{eq_wind_obc}
	\end{equation}
($W=0$ and $W=1$, respectively). The boundary between the trivial and topological phase is given by the curve $\left|\delta\right|=\sqrt{t^2-t_2^2}$, which simplifies to $t_2=\pm t$ when the system is Hermitian ($\delta$=0).
	
	\begin{figure*}[!htb]
%		\centering
		\hspace*{-1cm}     
		\includegraphics[width=1.0\linewidth]{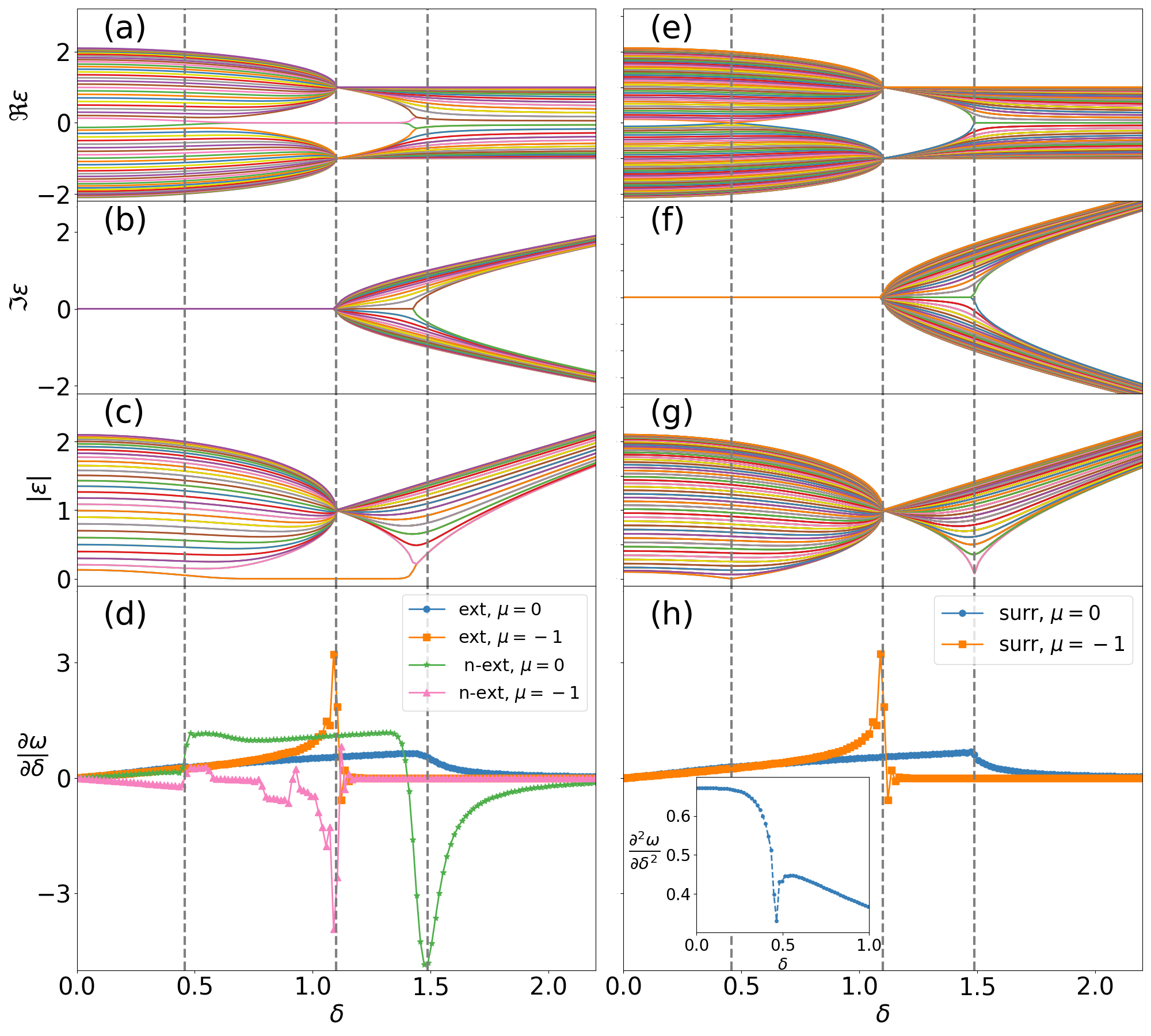}
		\caption{Comparison between the spectrum and grand potential of the system with OBC and the surrogate Hamiltonian for $t=1.1$.  (a) Real, (b) imaginary and (c) absolute values of energy for OBC. (d) First derivative of the grand potential: extensive part for $\mu=0$ (blue circles), extensive part for $\mu=-1.0$ (orange squares), non-extensive part for $\mu=0$ (green stars) and non-extensive part for $\mu=-1$ (pink triangles) for OBC. (e) Real, (f) imaginary and (g) absolute values of energy for the surrogate Hamiltonian. (h) First derivative of the grand potential for the surrogate Hamiltonian for $\mu=0$ (blue circles) and $\mu=-1.0$ (orange squares). The OBC simulations were done with $L=30$ unit cells, while for the surrogate Hamiltonian $100$ values of $k$ were used for the simulation of the spectrum and $1000$ values were used for the grand potential. The gray vertical dashed lines indicate the (theoretical) critical values of $\delta$.}
		\label{fig_obc}
	\end{figure*}
	
	For $\left|\delta\right|=t$, the non-Bloch band collapse occurs. For $\left|\delta\right|>t$, the energy, which was previously confined to the real axis, extends into the imaginary axis \cite{lee2019anatomy} [see Fig.~\ref{fig_obc} (b)]. The phases are then no longer deformable into their Hermitian counterparts, and we call these phases non-Hermitian topological and non-Hermitian trivial. The topology of theses phases is characterized in the generalized Brillouin zone, where the skin accumulation is ``unraveled'', and can be quantified by a winding number defined on the generalized Brillouin zone~\cite{yao2018edge,lee2019anatomy,lee2019unraveling}. The boundary between these phases is given by the curve $\left|\delta\right|=\sqrt{t_2^2+t^2}$.

%	 Although there is an extensive number of modes at the edge, these "bulk" modes with OBC can be obtained from the usual non-Bloch-bands by making an analytical continuation in $k$ to complex values~\cite{lee2019anatomy}. In the limit of $\Im k\rightarrow \infty$, the modes are exponentially localized at the edge of the system. Hence, some insight can be obtained in the topology of the system by doing a spectral flow in $k$~\cite{lee2019anatomy}. We combine this result with the wavefunction and illustrate the different phases of the open-boundary conditions system in Fig.~\ref{fig_phases_0.5_obc} for $t=0.5$ and in Fig.~\ref{fig_phases_2.0_obc} for $t=2.0$. It is clear from the phase diagram that the critical values $\delta_c$ are not the same as for PBC.

	\subsection{Thermodynamics}\label{subsec_thermo_ob}
	We now investigate how the spectrum of the system evolves as we vary $\delta$, such as to uncover unconventional critical thermodynamic behavior with no Hermitian analog. The results are shown in Fig.~\ref{fig_obc} (a)-(c) for $t=1.1$. Most strikingly, we observe the appearance of an asymmetric bunching of bands at the so-called non-Bloch-band collapse point~\cite{kunst2019non, alvarez2018non, longhi2020non, longhi2020_nonbloch, lee2019unraveling} $\delta_c=1.1$ (for all components), which is akin to a non-Hermitian flat band in complex energy space. As all bands can intermix with divergent density of states, such a point, despite not presenting a gap closing, is a phase transition point leading to band metric discontinuities linked to the NHSE-induced non-locality of the system~\cite{lee2019unraveling,lee2017band}. In addition, there are also topological transitions marked by the appearance/disappearance of zero modes, between the Hermitian topological and Hermitian trivial phase at $\delta_c=\sqrt{1.1^2-1}\approx 0.46$ and between the non-Hermitian topological to non-Hermitian trivial phase at $\delta_c=\sqrt{1.1^2+1}\approx 1.49$.

	As the non-Bloch band collapse is a gapped transition with criticality occurring at nonzero $\epsilon$, we need to set the chemical potential to that finite value to observe its signature in the grand potential. Fig.~\ref{fig_obc} (d) shows that indeed the transition between different Hermitian phases or between different non-Hermitian phases occurs for $\mu=0$, while the transition between the Hermitian and non-Hermitian topological phases occurs at $\mu=-1.0$ \footnote{Notice that the OBC phase transitions are a bit displaced with respect to the critical values of $\delta$. This is visible in the spectrum and in the grand potential derivative, and is a finite-size effect. For the surrogate Hamiltonian, instead, we can achieve the thermodynamic limit and the phase transition at $\delta_c$ is much sharper.}.

%	 We can determine these phase transitions using the grand potential, but to do that, we need to use a finite chemical potential, as the non-Bloch-band collapses at $\epsilon=-1$. To visualize these phase transitions, we then need to set $\mu$ to $0$ and $-1$, respectively. The results are presented in . For $\mu=-1.0$, there are signatures in the real and imaginary parts of the grand potential when there is a non-Bloch-band collapse. 
	 Interestingly, all the phase transitions show very distinctive behavior in the derivative of $\omega$ with respect to $\delta$. For the transition at $\delta_c\approx 0.46$, there is a discontinuity in the non-extensive contribution, while the extensive contribution shows a small kink. This imply that the second derivative presents a discontinuity, as can be seen in the inset of Fig.~\ref{fig_obc} (h). This suggest that they have the same scaling as the Hermitian SSH model (this will be later confirmed by the scaling analysis in Subsection \ref{subsec_scaling_ob}). On the other hand, for the transition at $\delta_c\approx 1.49$, the non-extensive part shows a discontinuity and the extensive part of $\omega$ also shows a kink\footnote{Notice that in the non-Hermitian trivial phase, the non-extensive contribution is nonzero although this phase does not present zero modes. This happens due to finite size effects, but this contribution vanishes for higher $\delta$.}. However, as in the case for PBC, the order of the phase transition cannot be unequivocally determined by only considering the grand potential derivative. For the non-Bloch band collapse at $\delta_c=1.1$, both contributions show a very similar behavior. This indicates that they might have the same order, which seems to contradict the Josephson hyperscaling relation. These aspects will be discussed in Section \ref{sec_nBBC}.
%	\begin{figure}[!htb]
%		\centering     
%		\includegraphics[width=1.0\linewidth]{surr_1.1_1000.png}
%		\caption{Spectrum of the surrogate Hamiltonian with $t=1.1$ for the (a) real, (b) imaginary and (c) absolute values of energy. (d) first derivative of the grand potential of this system for $\mu=0$ (blue circle) and $\mu=-1$ (orange square). Inset: second derivative of the grand potential close to $\delta_c\approx 0.46$. This simulation was done with $1000$ k values.}
%		\label{fig_surr}
%	\end{figure}   
	
	\subsection{Surrogate Hamiltonian}\label{subsec_surr}

	Due to the NHSE, the system with OBC has skin modes that diverge in the thermodynamic limit, making the exact determination of the thermodynamic behavior a very subtle question. A way to overcome this is to perform a change of basis that is position-dependent, which ``gauges away'' the skin mode accumulation, while preserving the OBC spectrum. Given an original Hamiltonian $H(k)$, one performs a complex deformation of the momentum $k\rightarrow k+i\kappa$, such that one obtains the so called ``surrogate Hamiltonian" $H(k+i\kappa)$, which does not experience the NHSE~\cite{lee2019unraveling, helbig2020generalized} (by construction, a constant complex momentum deformation does not change the OBC spectrum~\cite{lee2019anatomy}). For the non-Hermitian SSH Hamiltonian~\eqref{eq_bloch_ham_nh_ssh}, the deformation is~\cite{yao2018edge}
	\begin{equation}
		\kappa=-  \log\left(\sqrt{\left|\frac{t-\delta}{t+\delta}\right|}\right),
		\label{eq_k_complex}
	\end{equation} 
such that the Bloch Hamiltonian of Eq.~\eqref{eq_bloch_ham_nh_ssh} becomes 
	\begin{equation}
	\begin{split}
		H\left(k\right)=&\left(t_1-\delta+e^{i\left(k+i \kappa\right)}\right)\sigma_{+}\\
						&+\left(t_1+\delta+e^{-i\left(k+i \kappa\right)}\right)\sigma_{-}
		\label{eq_H_surr_bloch},
	\end{split}
	\end{equation}
where $\sigma_{\pm}=\left(\sigma_x\pm i\sigma_y\right)/2$. As $\sigma_x\sigma_{\pm}\sigma_x=\sigma_{\mp}$, the similarity transformation $U$ in Eq.~\eqref{eq_PH} is given simply by $\sigma_x$ for this model. 

	The spectrum of this system,
	\begin{equation}	
		\epsilon_\pm\left(k\right)=\pm \sqrt{1^2+t^2-\delta^2+2\sqrt{t^2-\delta^2}\cos k},
		\label{eq_epsilon_kappa}
	\end{equation}
 is identical to the spectrum of the original model under OBC~\cite{lee2019unraveling, kawabata2019symmetry}, with the exception of the zero modes [compare Fig.~\ref{fig_obc} (a)-(c) with Fig.~\ref{fig_obc} (e)-(g)]. This is a way to recover the bulk-boundary correspondence. %Furthermore, this system does not present the NHSE, what makes it stable in the thermodynamic limit.

	Theses modes are the ones associated to the extensive part of the system with OBC. Therefore, one can obtain the extensive contribution of the grand potential from the surrogate Hamiltonian, as it can be observed comparing Fig.~\ref{fig_obc} (d) and Fig.~\ref{fig_obc} (h). The advantage of using the surrogate Hamiltonian is that we can approach the thermodynamic limit and make the thermodynamic results more reliable because it is not unstable for large system sizes. It is more evident then that indeed the first derivative of $\omega_\text{ext}$ shows a discontinuity at the non-Bloch band collapse $\delta_c=1.1$, while it shows a kink at $\delta_c\approx 1.49$. As the behavior at $\delta_c\approx 0.49$ is not clear, we calculate the second derivative around this point [inset of Fig.~\ref{fig_obc} (h)]. It shows a discontinuity, indicating that there is a kink in the first derivative.
	
%	 Had the original Hamiltonian, which is susceptible to the NHSE, been studied instead of the surrogate Hamiltonian, topological and all other skin states will all be pushed to the boundary and be harder to thermodynamically extricate. 
	
	\begin{figure}[!htb]
		\centering
		\includegraphics[width=\linewidth]{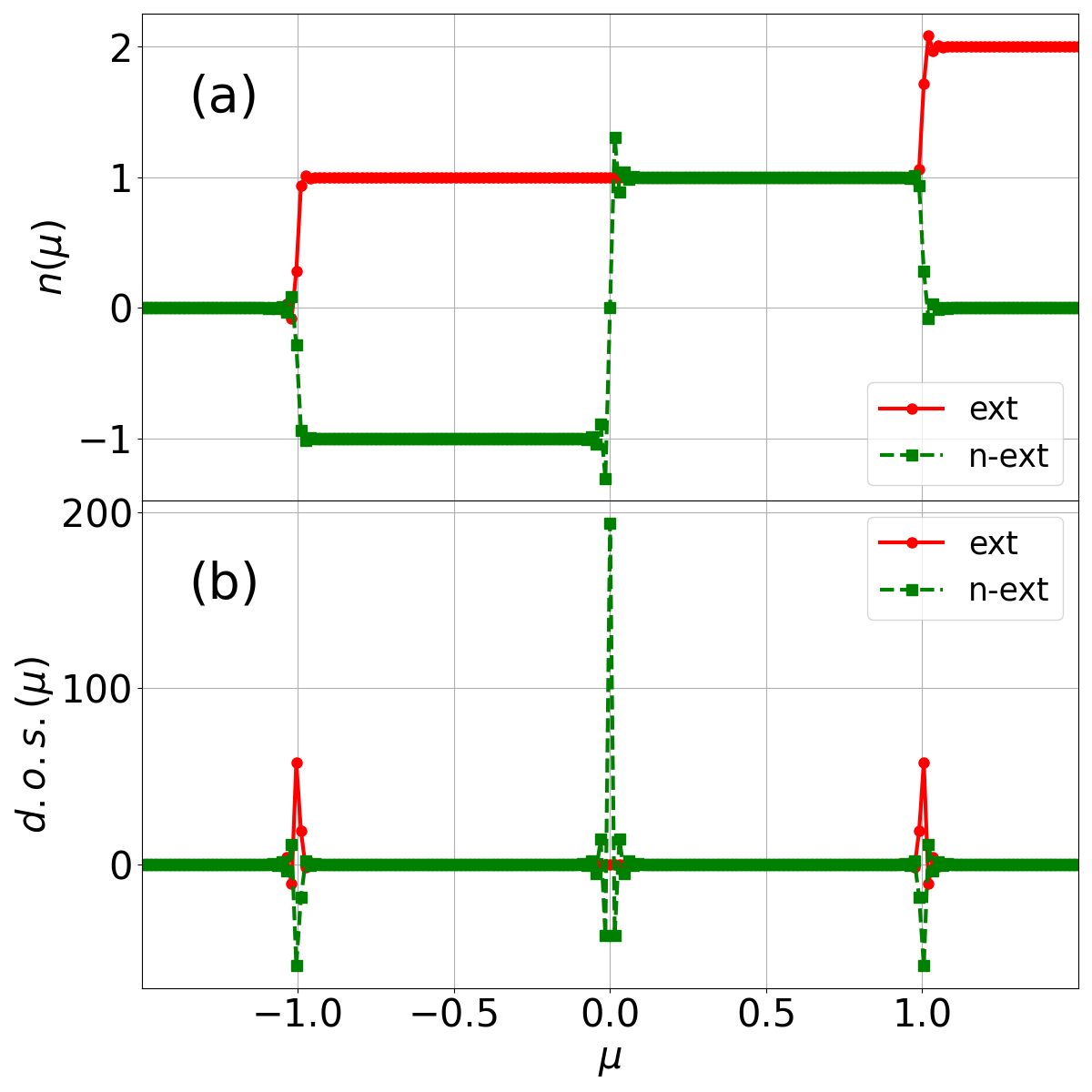}
		\caption{(a) The electronic density  $n\left(\mu\right)$ and (b) the reduced density of states $d.o.s(\mu)$ for the extensive (red circles) and non-extensive (green squares) contributions as a function of $\mu$ for $t=1.1$ and $\delta=1.1$, where the non-Bloch band collapse occurs. We used $50$ unit cells in these simulations.}
		\label{fig_dens_nBBC}
	\end{figure}
\subsection{Electronic density and density of states} 
  	We next examine the divergent degeneracy at the non-Bloch band collapse~\cite{lee2019unraveling, alvarez2018non, longhi2020non, longhi2020_nonbloch} through the electron density $n$, as derived from the rate of change of $\omega$ with $\mu$
  	\begin{equation}
  		n\left(\mu\right)=\frac{N}{L}=-\frac{\partial \omega}{\partial \mu},
  	\end{equation}
where $N$ is the number of electrons, and through the (reduced) density of states $d. o. s. $, which can be obtained by the second derivative of $\omega$ with respect to $\mu$,
	\begin{equation}
		d. o. s. \left(\mu\right)=-\frac{\partial^2\omega}{\partial \mu^2}.
	\end{equation}
	The results for all phases are described in Appendix \ref{app_dens}, but if we focus on $\delta=1.1$, where the non-Bloch band collapse occurs, we see in Fig.~\ref{fig_dens_nBBC} that the electronic density shows only steps of integer occupancy, depending on the chemical potential [Fig.~\ref{fig_dens_nBBC} (a)] contributions! This leads to the van Hove singularities seen in Fig.~\ref{fig_dens_nBBC} (b), and indicates the remarkable character of this transition, where there is an accumulation of modes in a fermionic system and the density of states is composed only of van Hove singularities~\cite{alvarez2018non,longhi2020non, longhi2020_nonbloch}.

	Another very intriguing feature observed both in the electronic density and in the density of states is that, at the non-Bloch collapse, the extensive and non-extensive contributions are exactly opposite to each other (for the electronic density at $\mu=1$ with respect to the $n=1$ line). This can be understood if we assume that the non-extensive modes can act like a reservoir to the extensive modes. This was observed in a heat-machine approach to the topological mode of the Kitaev model \cite{yunt2019topological}, where the edge mode acted as a heat reservoir to the bulk modes. Here, it seems that this is the case but now the non-extensive modes seems to act as a particle reservoir for the extensive modes. A similar feature can be observed for all phases (see Appendix \ref{app_dens}), although they are not exactly opposite to each other.

\subsection{Critical Exponents}\label{subsec_scaling_ob}

	In principle, we can use the surrogate Hamiltonian to obtain the critical exponents $z$ and $\nu$ and compare them to the result for the scaling of $\omega$, as it was done for PBC. From the expression of the spectrum in Eq.~\eqref{eq_epsilon_kappa}, it is clear that the gap closing occurs for:
	\begin{equation}
	\begin{cases}
		\delta_c=\sqrt{t^2-1} \qquad K_c=\pi \qquad \mu=0,\\
		\delta_c=\sqrt{t^2+1} \qquad K_c=\pi/2 \qquad \mu=0,\\
		\delta_c=t \qquad \forall k \qquad \mu=\pm 1.
	\end{cases}
		\label{eq_ob_crit_points}
	\end{equation}
	For the transitions that happen at $\delta_c=\sqrt{t^2\pm 1}$, the gap closing can be described using the same kind of calculation done for PBC, with critical exponents
	\begin{equation}
	\begin{cases}
		\delta_c=\sqrt{t^2-1} \qquad z=1 \qquad \nu=1,\\
		\delta_c=\sqrt{t^2+1} \qquad z=1/2 \qquad \nu=1,
	\end{cases}
	\end{equation}
as it can be seen in Figs.~\ref{fig_scaling_surr} (a) and (b), respectively (see Appendix \ref{app_gap} for details of the calculation). This makes that the grand potential scales with $g^{1}$ (both for the extensive and non-extensive components due to $\xi$, see discussion on Subsection \ref{subsec_scaling}) at $\delta_c=\sqrt{t^2-1}$, and as $g^{1/2}$ (again, for both components) for $\delta_c=\sqrt{t^2+1}$. This is indeed confirmed in Fig.~\ref{fig_scaling_surr} (c), where we calculated $\omega_s$ using Eq.~\eqref{eq_omega_sing} for both values of $\delta_c$.

	\begin{figure}[!htb]
		\centering
		\includegraphics[width=\linewidth]{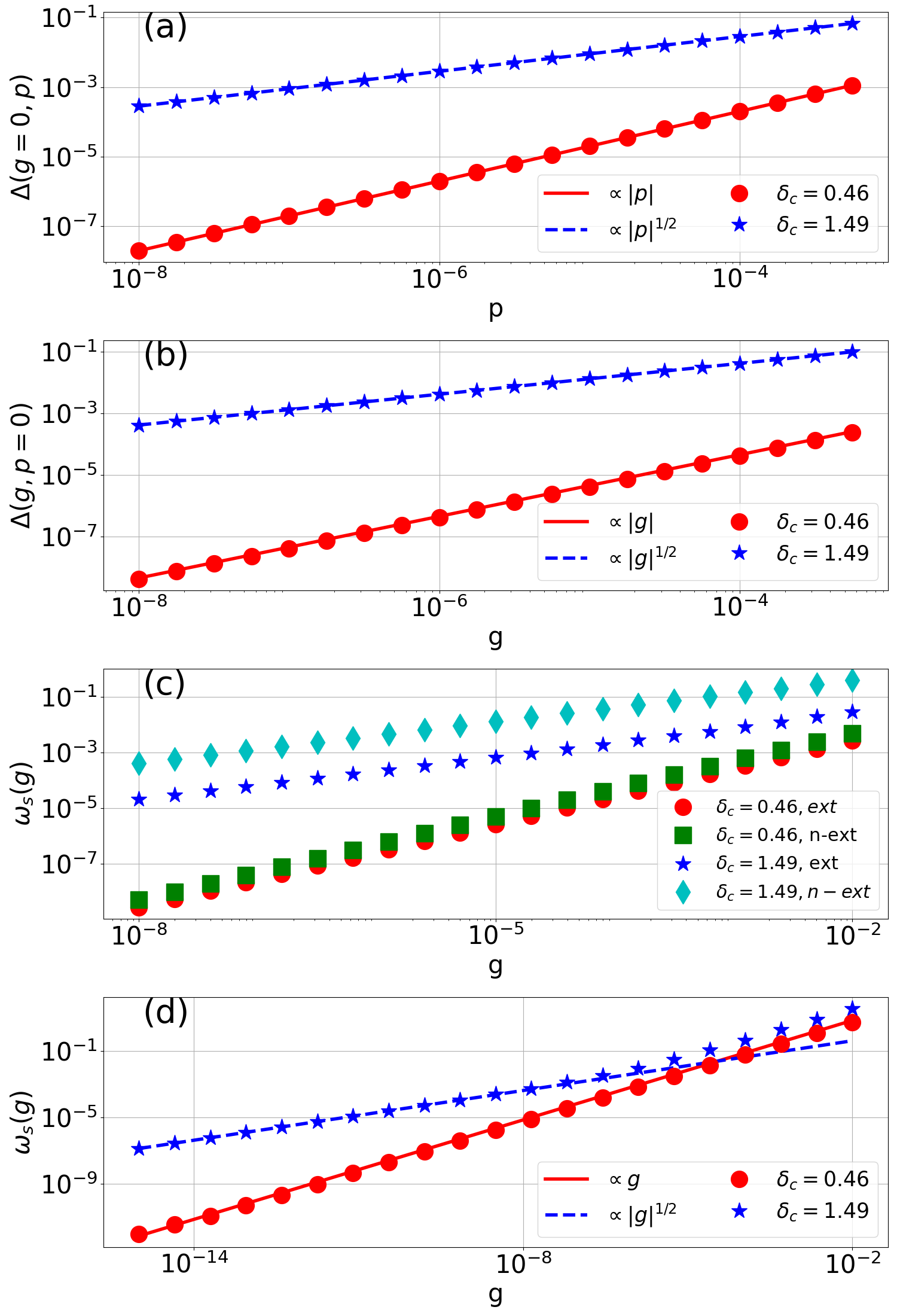}
		\caption{Scaling analysis for the gap $\Delta$ closing and for the singular part of the grand potential $\omega_s$ near the critical points $\delta_c\approx 0.46$, $K_c=\pi$ and $\delta_c\approx 1.49$, $K_c=\pi/2$ for $t=1.1$. (a) Gap closing in momentum space, for both phase transitions, compared to the curves $\left|p\right|$ (blue dashed lines) and $\left|p\right|^{1/2}$ (red solid lines). (b) Closing of the gap in the parameter space for both phase transitions, compared to the curves $\left|g\right|$ (blue dashed lines) and $\left|g\right|^{1/2}$ (red solid lines). (c) Scaling of $\omega_s$ for both phase transitions and both components calculated for $L=20$ and open boundary conditions. (d) Scaling of $\omega$ calculated from the surrogate Hamiltonian for both phase transitions and $1000$ $k$ values.}
		\label{fig_scaling_surr}
	\end{figure}

	Although the result for the first transition is expected, as it is the transition between the ``Hermitian" phases and should have the same critical exponents even with finite $\delta$, the fact that the transition between the non-Hermitian phases have particular critical exponents and that they are the same as the one observed for PBC is really interesting. In addition, if we increase the system size using the surrogate Hamiltonian spectrum, as done in Fig.~\ref{fig_scaling_surr} (d), we see that the $\delta_c=\sqrt{1+t^2}$ transition also exhibits a change in the scaling behavior for increasing values of $g$ (as seen for PBC), while for the other transition it maintains the $g^{1}$ scaling. 

\subsubsection{Hermitian and non-Hermitian Dirac universality classes}

	The results for the critical exponents can be understood if one considers that the Hermitian SSH model pertain to the universality class of the Dirac model \cite{chen2017correlation}, while the transition between non-Hermitian phases belongs to the universality class of the non-Hermitian Dirac model \cite{shen2018topological, rui2019topology, zirnstein2019bulk}.
	
	The Hermitian Dirac model has the Bloch Hamiltonian
	\begin{equation}
		h(p)=M\sigma_{x}+v_F p \sigma_y,
		\label{eq_herm_Dirac}
	\end{equation}
where the mass term $M$ tunes the phase transition between the trivial ($M>0$) and topological phase ($M<0$) and $v_F$ is the Fermi velocity.

	Consider the Hermitian SSH model,
	\begin{equation} 
		h(p)=\left[t+t_2\cos(k)\right]\sigma_{x}+t_2\sin(k) \sigma_y,
		\label{eq_herm_SSH}
	\end{equation}
in the vicinity of the topological phase transition at $t=t_2$ and $K_c=\pi$, such that we can write $t=t_2\left(1+M\right)$ and $k=K_c+p$. We then obtain
	\begin{equation} 
		h(p)\approx M\sigma_{x}-t_2 p \sigma_y,
		\label{eq_herm_SSH_crit}
	\end{equation}
which has the form of Eq.~\eqref{eq_herm_Dirac} upon identifying $t_2$ with $-v_F$.

	A version of the non-Hermitian Dirac model \cite{shen2018topological, rui2019topology, zirnstein2019bulk},
	\begin{equation}
		h(p)=M\sigma_{x}+(v_F p+ i \gamma)\sigma_y,
		\label{eq_n_herm_Dirac}
	\end{equation} 
presents a new phase transition when $\gamma=\pm M$, in addition to the Hermitian one at $M=0$ and $\gamma=0$.

	If we consider the non-Hermitian SSH model in Eq.~\eqref{eq_bloch_ham_nh_ssh} with $t=t_2\left(1+M\right)$ and $k=K_c+p$, we get  
	\begin{equation} 
		h(p)\approx M\sigma_{x}+(-t_2 p+i \delta) \sigma_y,
		\label{eq_herm_nh_SSH_crit}
	\end{equation}
such that it is mapped on the non-Hermitian Dirac model in Eq.~\eqref{eq_n_herm_Dirac} if we identify $t_2$ with $-v_F$ and $\delta$ with $\gamma$. 

	\section{Scaling for the non-Bloch band collapse}\label{sec_nBBC}

	For the non-Bloch band collapse ($\delta_c=t$), the energy $\epsilon$ given by Eq.~\ref{eq_epsilon_kappa}, is equal to $\pm 1$ for any value of $k$, such that the gap closes in a flat band. Hence, although the gap scales with $g^{1/2}$, as calculated in Appendix \ref{app_gap}, it is difficult to obtain the critical exponent $z$ using the scaling of the gap closing in momentum space. However, as the exponent $\nu$ defines how the correlation length diverges at to the phase transition
	\begin{equation}
		\xi=\left|g\right|^{-\nu},
		\label{eq_scaling_xi}
	\end{equation}
one can obtain $\nu$ by performing a scaling analysis of the correlation length.

	To obtain $\xi$  from the simulation, we assume that $\left|\langle a_{0}^{\dagger}a_{r}\rangle\right|=\left|\langle a_{0}^{\dagger}a_{0}\rangle\right|\exp\left(-r/\xi\right)$, where $0$ denotes the first site, such that a linear fit of the form
	\begin{equation}
		\log\left(\frac{\left|\langle a_{0}^{\dagger}a_{r}\rangle\right|}{\left|\langle a_{0}^{\dagger}a_{0}\rangle\right|}\right)=-\frac{r}{\xi},
		\label{eq_xi}
	\end{equation}
allows one to obtain $\xi$.
	
	As discussed in Section \ref{sec_thermo}, for non-Hermitian systems there are two different wavefunctions $\ket{\psi^{L}}$ and $\ket{\psi^{R}}$, such that they can lead to different correlation functions. Then, one can use different definitions to evaluate these correlation functions, as detailed in Appendix \ref{app_corr}. In Fig.~\ref{fig_crit_nBBC} (a), we show the results for the correlation functions (at $T=0$ and $\mu=-1$) calculated using $\ket{\psi^{L}}$ as a basis (denoted by $\Braket{LL}$) and using $\bra{\psi^{L}}$ and $\ket{\psi^{R}}$ (denoted by $\Braket{LR}$) . Surprisingly, the correlation lenght does not seem to diverge as we approach the critical point for none of the correlation functions.

	One possible explanation is that the NHSE introduces a new length scale, given by the skin depth $\kappa^{-1}=\left[\log\sqrt{\left|\left(t+\delta\right)/\left(t-\delta\right)\right|}\right]^{-1}$, which invalidates scaling arguments based on a single correlation length scale. $\kappa^{-1}$ goes to zero as $g$ goes to zero, which makes the correlation lenght defined in Eq.~\eqref{eq_xi} goes to zero as $g$ tends to zero! The fact that the correlation length does not diverge in this phase transition are supported in Fig.~\ref{fig_crit_nBBC} (a) both for the simulated data and for the theoretical expression for $\Im k^{-1}$ obtained from Eq.~\eqref{eq_k_complex}. 

	\begin{figure}[!htb]
		\centering
		\includegraphics[width=\linewidth]{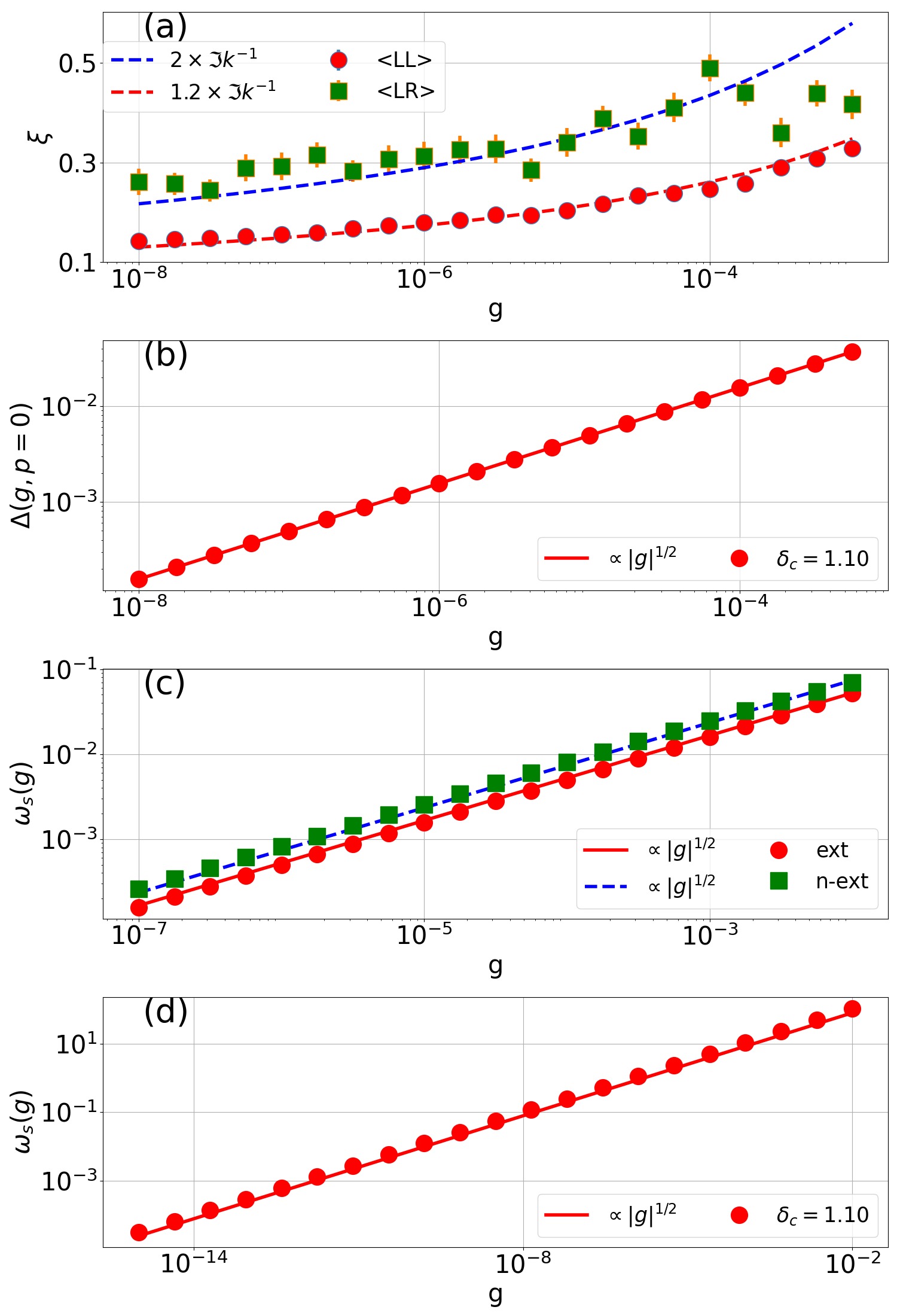}
		\caption{Scaling analysis for the correlation lenght $\xi$, gap $\Delta$ closing and singular part of the grand potential $\omega_s$ near the critical points $\delta_c=1.1$, $K_c=0$ for $t=1.1$. (a) Correlation length as a function of $g$ close to the phase transition for the correlation lenght using only left eigenvectors ($<LL>$, red circles) and using left  and right eigenvectors ($<LR>$, green squares) compared to curves of the inverse of $\kappa$ (dashed lines with different amplitudes) defined in Eq.~\eqref{eq_k_complex}. (b) Closing of the gap in the parameter space (red circles), compared to the curve  $\left|g\right|^{1/2}$ (red solid line). (c) Scaling of $\omega_s$ for the extensive (red circles) and non-extensive (green squares) compared to curves proportional to $g^{1/2}$ (red solid line and blue dashed line) calculated for $L=20$ and open boundary conditions compared to . (d) Scaling of $\omega$ calculated from the surrogate Hamiltonian with $1000$ $k$ values (red circles) compared to a curve proportional to $g^{1/2}$ (red solid line).}
		\label{fig_crit_nBBC}
	\end{figure}
	
	This contradicts the usual lore of phase transitions, as in a phase transition the divergence of the correlation length indicates that the system is very sensitive to fluctuations (classical or quantum), which can drive the phase transition. This is caused by the very peculiar character of the non-Bloch band collapse: as all modes are located at the edge (even for the surrogate Hamiltonian for $\delta=t$), so that the system is very unstable. This instability is manifest on the change of the branch of energies, which is similar to a change in ground state. In addition, the fact that the gap closes as a flat band shows that the spatial and time fluctuations are decoupled and there is no typical length scale associated to the gap closing and change of ground state.

	The fact that there is still a phase transition is clear when a scaling analysis is performed in the gap and in the grand potential as a function of $g$, see Fig.~\ref{fig_crit_nBBC} (b)-(d). Although the critical exponent $\nu$ is not defined in this phase transition, the way in which the gap closes in the parameter space suggests that $\nu z$ is equal to $1/2$. As $\xi$ does not have a characteristic scaling, this imply that the order of the phase transition should not depend on the dimension of the system, such that the Josepshon hyperscaling relation is not satisfied. This implies that both components scale with $g^{1/2}$, which is consistent with the result obtained in Fig.~\ref{fig_obc} (d), where both the extensive and non-extensive contributions show basically the same behavior. This establishes the complete exceptional character of the non-Bloch band collapse as a quantum phase transition. 

\section{Conclusions}\label{sec_conc}

	We developed a framework for characterizing phase transitions in non-Hermitian systems. Despite the fact that non-Hermitian systems exhibit complex energies, for models with pseudo-Hermitian symmetry, the grand potential is a real quantity, such that the traditional thermodynamics analysis can be applied. 
	
	As non-Hermitian systems present in general different phase diagrams for different boundary conditions, we analyzed separately the case for PBC and OBC. We also investigated system a ``Hermintianized" version of the model and the so called surrogate Hamiltonian, which yields further insight into the phases of these models. Most phase transitions in this system can be characterized using critical exponents and belong to either the Dirac or non-Hermitian Dirac universality class.
	
	However, the non-Bloch band collapse transition possesses an additional length scale (skin depth) and does not strictly obey critical scaling relations. This makes that the correlation length does not diverge, but rather goes to zero at this phase transition. This very peculiar behavior is due to the localization of the modes at the edges due to the complexification of the momentum. This absence of divergence in the correlation length breaks the Josephson hyperscaling relation. This implies that the order of the phase transition is equal for the extensive and non-extensive parts of this system, although they have different dimensions. The breaking of the Josepshon hyperscaling relation marks the extremely exceptional character of phase transitions that are possible in non-Hermitian systems.  

	The anomalous scaling laws discussed here could be possibly observed in cold-atom systems, which are a promising platform to observe the NHSE \cite{li2020topological}. Nevertheless, this would require an extension of our analysis to finite temperatures. The generalization of this formalism for other non-Hermitian systems, such as the ones that are not pseudo-Hermitian, or the ones that have higher dimension and present multiple scaling of the NHSE are natural extensions of this work. 
	
\section*{Acknowledgements}

	The authors thank Natanael de Carvalho Costa and Flore Kunst for a careful reading of the manuscript. RA acknowledges funding from the Brazilian Coordination for the Improvement of Higher Education Personnel (CAPES) and from Delta Institute for Theoretical Physics (DITP) consortium, a program of the Netherlands Organization for Scientific Research (NWO) that is funded by the Dutch Ministry of Education, Culture and Science. CH acknowledges support from the Singapore MOE Tier I grant (WBS No. R-144-000-435-133).
%\clearpage

\appendix

	\section{Scaling of $\Omega$ with $L$}\label{app_scaling_L}

\begin{figure}[!htb]
		\centering
		\includegraphics[width=\linewidth]{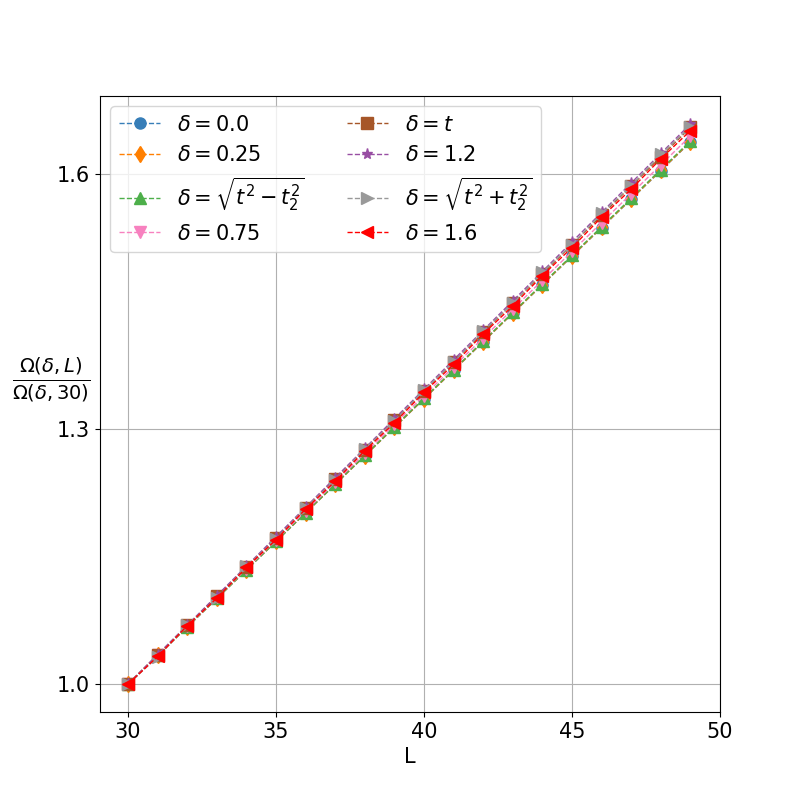}
		\caption{Grand potential as a function of $L$ for many representative values of $\delta$. The blue circles are results for the Hermitian model ($\delta=0$). The orange diamonds are results for the Hermitian trival phase ($\delta=0.25$). The green triangles are results for the transition between the Hermitian trivial and Hermitian non-trivial phase ($\delta=\sqrt{t^2-t_2^2}$). The pink reversed triangle are results for the Hermitian non-trivial phase ($\delta=0.75$). The brown squares are results for the region in the proximity of the non-Bloch band collapse ($\delta=t$). The purple stars are results for the non-Hermitian trivial phase ($\delta=1.2$). The gray triangle pointing to the right are results for the transition between the non-Hermitian trivial and the non-Hermitian topological phase ($\delta=\sqrt{t^2+t_2^2}$). The red triangles pointing to the left show results for the non-Hermitian topological phase ($\delta=1.6$). We used $t=1.1$ and $\mu=0$ for all the simulations. We divide the result for each $\delta$ by their value at $L=30$, such that we can see the scaling for all values of $\delta$. }
		\label{fig_scaling_L}
	\end{figure}

	Although non-Hermitian systems present many interesting scalings with $L$ due to their non-local character, we verified that Eq.~\eqref{eq_Omega_L} always holds for this model . In Fig.~\ref{fig_scaling_L}, we show that for all the phases arising in the phase diagram for OBC, the grand potential scales linearly with $L$. However, at precisely the non-Bloch band collapse point, $\Omega\left(\delta=t, \mu=-1\right)=0$, as can be anticipated by the spectrum in Fig.~\ref{fig_obc}. Thus the grand potential does not scale with the system size at this point.

\section{Density of states for all phases}\label{app_dens}
	\begin{figure*}[!htb]
		\centering
		\includegraphics[width=\linewidth]{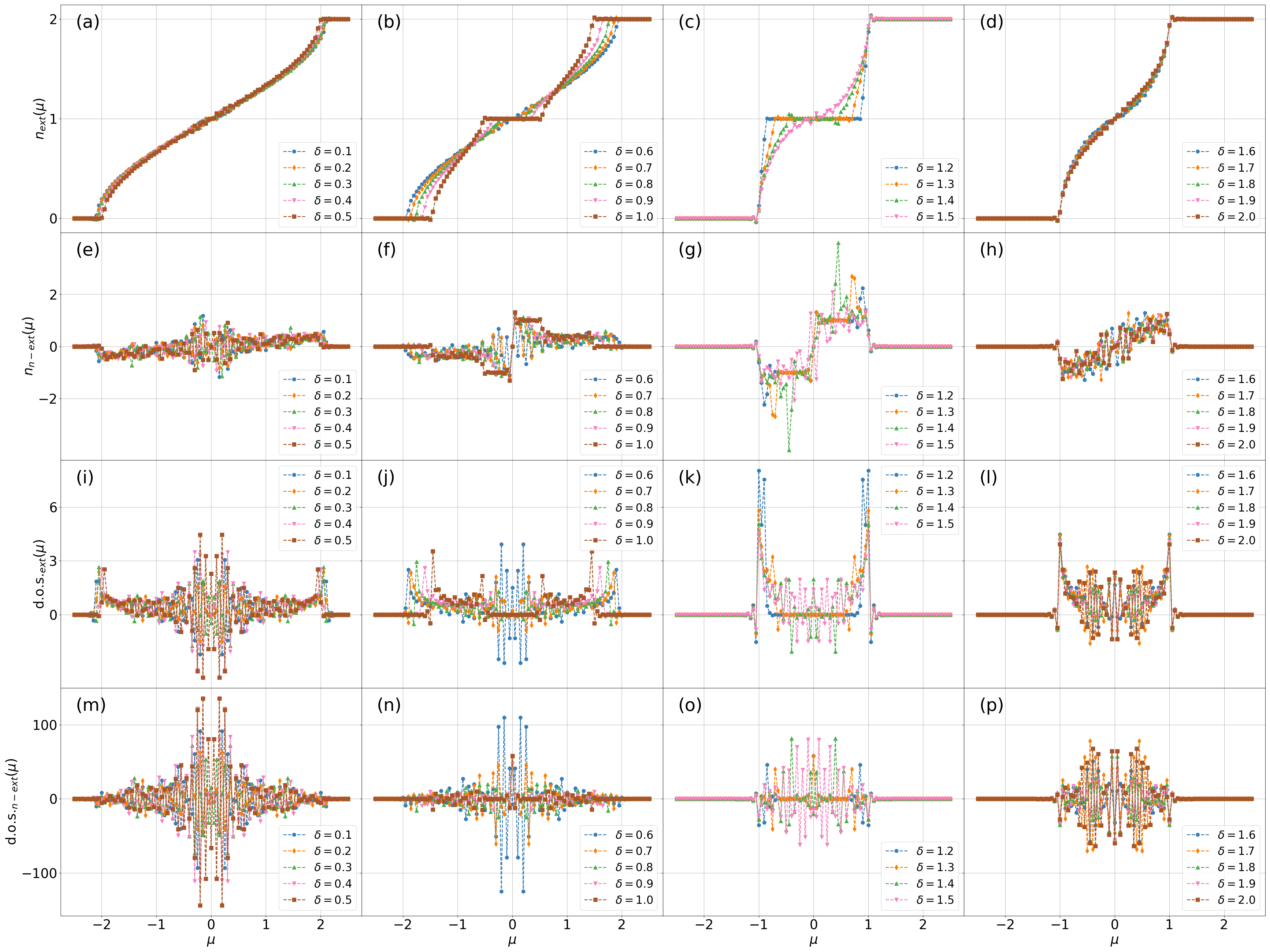}
		\caption{Electronic density $n$ and d.o.s. as a function of the chemical potential $\mu$ for all phases and for both components of the grand potential, for $t=1.1$ and several values of $\delta$. Electronic density for the extensive contribution in the (a) Hermitian trivial, (b) Hermitian topological, (c) non-Hermitian topological and (d) non-Hermitian trivial phases. Electronic density for the non-extensive contribution in the (e) Hermitian trivial, (f) Hermitian topological, (g) non-Hermitian topological and (h) non-Hermitian trivial phases. Reduced density of states for the extensive contribution in the (i) Hermitian trivial, (j) Hermitian topological, (k) non-Hermitian topological and (l) non-Hermitian trivial phases. Reduced density of states for the non-extensive contribution in the (m) Hermitian trivial, (n) Hermitian topological, (o) non-Hermitian topological and (p) non-Hermitian trivial phases. We used $50$ unit cells in these simulations.}
		\label{fig_dens_phases}
	\end{figure*} 

	The results for the electronic density and reduced density of states for all phases are presented in Fig.~\ref{fig_dens_phases} for $t=1.1$. We can understand better these results by comparing them to the spectrum in Fig.~\ref{fig_obc}. For the Hermitian trivial phase, Figs.~\ref{fig_dens_phases} (a), (e), (i) and (m), the spectrum presents a very small gap such that the results are similar to a metallic state: there is a continuous increase of the electronic density as we increase the chemical potential for the extensive component, while the non-extensive component shows a very small, but finite, density. The d.o.s reflects this result, as it is finite for all energy values inside the band, but does not show any distinct feature.
	
	For the Hermitian topological phase, the gap increases and the bands become less dispersive, what makes that the electronic density for the extensive contribution show zero density of electrons, until it reaches some value of $\mu$ that is inside the continuum of bands [see Fig.~\ref{fig_dens_phases} (b)]. Then, the density states increase until it reaches the top limit of this continuum and displays a plateau at half-filling. These are also present in the density of states, Fig.~\ref{fig_dens_phases} (j), which signals the gap (although with much fluctuations). Besides the features seen in the extensive contribution, the non-extensive contribution of the electronic density, see Fig.~\ref{fig_dens_phases} (f), shows a jump at $\mu=0$ due to the zero mode, what is reflected in the van Hove singularity in the density of states for $\mu=0$, see Fig.~\ref{fig_dens_phases} (n).
	
	The same features are visible in the non-Hermitian topological phases [Fig.~\ref{fig_dens_phases} (c), (g), (k), (o)], except that they are amplified because the bands are less dispersive. For the non-Hermitian trivial phase [Fig.~\ref{fig_dens_phases} (d), (h), (l), (p)], we see the same features as for the Hermitian trivial phase, but again amplified due to the less dispersive character of the bands.

\begin{widetext}
\section{Gap closing for open boundary conditions}\label{app_gap}
	
	We can do the same kind of calculation for the critical points $\delta_c=\sqrt{t^2\pm 1}$ of the surrogate spectrum in Eq.~\eqref{eq_epsilon_kappa}, as it was done in Subsection \ref{subsec_scaling} for the periodic case.
	
	Starting with the critical point $\delta_c=\sqrt{t^2-1}$, $K_c=\pi$, we find

	\begin{align}
	\begin{split}
		\epsilon(g=0, p)\approx \sqrt{1+t^2-(t^2-1)+2\sqrt{1}\left(-1+\frac{p^2}{2}\right)}=p,
	\end{split}
	\end{align}
implying that $z=1$. In addition,
	\begin{align}
	\begin{split}
		\epsilon(g, p=0)\approx &\sqrt{1+t^2-(t^2-1)(1+g)^2-2\sqrt{t^2-(t^2-1)(1+g)^2}}\\
		\approx&\sqrt{2-(t^2-1)(2g+g^2)-2+2(t^2-1)g+(t^4-t^2)g^2}\\
		=&\sqrt{t^2-1}t \ g,
	\end{split}
	\end{align}
hence $\nu z=1$ and $\nu=1$.

	For the critical point  $\delta_c=\sqrt{t^2+1}$, $K_c=\pi/2$, we have
%\begin{widetext}
	\begin{align}
	\begin{split}
		\epsilon(g=0, p)\approx \sqrt{1+t^2-(t^2+1)+2\sqrt{-1}p}=\frac{1+i  }{2}p^{1/2},
	\end{split}
	\end{align}
%\end{widetext}
implying that $z=1/2$, and
	\begin{align}
	\begin{split}
		\epsilon(g, p=0)\approx \sqrt{1+t^2-(1+t^2))\left(1+g\right)^2}=t\sqrt{-2g-g^2}\approx \sqrt{2}t\left|g\right|^{1/2}\sqrt{-\text{sgn}\left(g\right)},
	\end{split}
	\end{align}
which presents the same asymmetry as obtained for the PBC case, so that $\nu z=1/2$ and $\nu=1$. The critical exponents for both phase transitions are supported by the scaling results from the simulations in Figs.~\ref{fig_scaling_surr} (a) and (b).

	For the non-Bloch band collapse ($\delta_c=t$) there is no gap closing, but we can analyze how the energy approaches the flat band value ($\epsilon=1$). In this case:
	\begin{align}
	\begin{split}
		\epsilon(g, k)&= \sqrt{1+t^2-t^2\left(1+g\right)^2+2\sqrt{t^2-t^2\left(1+g\right)^2}\cos k}=\sqrt{1-t^2\left(2g+g^2\right)+2t \sqrt{-2g+g^2}\cos k}\\
		&\approx \sqrt{1+2\sqrt{2}t \sqrt{-g}\cos k}\\
		&\approx 1+\sqrt{2}t\cos k\sqrt{-\text{sgn}\left(g\right)} \ \left|g\right|^{1/2},
	\end{split}
	\end{align}
such that $\nu z=1/2$.
\end{widetext}

\section{Correlation functions for non-Hermitian systems}\label{app_corr}

	For a generic (non-interacting) tight-binding model, we can write the second quantized Hamiltonian in terms of a matrix $\mathbb{H}$,
	\begin{equation}
		H=\sum\limits_{\alpha,\beta}c^\dagger_{\alpha}h_{\alpha \beta}c_{\beta}=C^\dagger \mathbb{H} \ C,
	\end{equation}
where $\alpha$ and $\beta$ represent generic quantum numbers and $C=(\begin{matrix}c_{1}&c_{2}&\cdots\end{matrix})^T$ is a vector with all the annihiliation operators. 

	We can diagonalize the Hamiltonian to obtain
	\begin{equation}
		C^\dagger \mathbb{H} \ C=\Psi^\dagger \mathbb{E}\Psi,
	\end{equation}
where $\mathbb{E}=\mathbb{S}^{-1} \mathbb{H}\ \mathbb{S}$ is a matrix with the energies of the system in its diagonal and $\Psi=\left(\begin{matrix}\psi_{1}&\psi_{2}&\cdots\end{matrix}\right)^T$ is a vector of the annihiliation operators of the eigenmodes of the Hamiltonian. $\mathbb{S}$ relates $\Psi$ and $C$,
	\begin{equation}
		\Psi=\mathbb{S}^{-1} C\Rightarrow C=\mathbb{S} \Psi
	\end{equation}
which leads to
	\begin{equation}
		c_{\alpha}=\sum\limits_{m}\mathbb{S}_{\alpha m} \psi_m,
	\end{equation} 
with $m$ labelling the energy mode. The correlation function for a Hermitian system is then given by
	\begin{equation}
		\braket{c_{\rho}^\dagger c_{\sigma}}(T, \mu)=\sum\limits_{m}\mathbb{S}^{*}_{\rho m}\mathbb{S}_{\sigma m}f_{FD}(\epsilon_m, \beta, \mu),
	\end{equation}
where $f_{FD}=\left(e^{\beta\left(\epsilon_m-\mu\right)}+1\right)^{-1}$ is the Fermi-Dirac distribution. For $T=0$
	\begin{equation}
		\braket{c_{\rho}^\dagger c_{\sigma}}(T=0, \mu)=\sum\limits_{\epsilon_m<\mu}\mathbb{S}^{*}_{\rho m}\mathbb{S}_{\sigma m}.
	\end{equation}
	
	For non-Hermitian systems, $H^\dagger$ and $H$ are different such that
	\begin{equation}
		C=\mathbb{S}^{R}\Psi^{R}=\mathbb{S}^{L}\Psi^{L},
	\end{equation}
where $\Psi^{R}$ and $\Psi^{L}$ are the set of eigenmodes of $H$ and $H^\dagger$, respectively.

	In this way, we have four kinds of correlation functions:
	\begin{equation}
	\begin{cases}
		\Braket{c^\dagger_\rho c_\sigma}^{RR}(T=0, \mu)=\sum\limits_{\Re\epsilon_m<\mu} \left(\mathbb{S}^{R}_{\rho m}\right)^*\mathbb{S}^{R}_{\sigma m}\\
		\Braket{c^\dagger_\rho c_\sigma}^{RL}(T=0, \mu)=\sum\limits_{\Re\epsilon_m<\mu} \left(\mathbb{S}^{R}_{\rho m}\right)^*\mathbb{S}^{L}_{\sigma m}\\
		\Braket{c^\dagger_\rho c_\sigma}^{LR}(T=0, \mu)=\sum\limits_{\Re\epsilon_m<\mu} \left(\mathbb{S}^{L}_{\rho m}\right)^*\mathbb{S}^{R}_{\sigma m}\\
		\Braket{c^\dagger_\rho c_\sigma}^{LL}(T=0, \mu)=\sum\limits_{\Re\epsilon_m<\mu} \left(\mathbb{S}^{L}_{\rho m}\right)^*\mathbb{S}^{L}_{\sigma m}.
	\end{cases}
	\end{equation}

	From the above expressions, it is clear that $\Braket{c^\dagger_\rho c_\sigma}^{RL}=\left(\Braket{c^\dagger_\sigma c_\rho}^{LR}\right)^*$, such that these correlations functions are not independent. Similarly, for pseudo-Hermitian systems, $\Braket{c^\dagger_\rho c_\sigma}^{RR}=\left(\Braket{c^\dagger_\sigma c_\rho}^{LL}\right)^*$.
	
	For the results in Section \ref{sec_nBBC}, we use the correlation functions $\Braket{a^\dagger_0 a_r}^{LL}$ and $\Braket{a^\dagger_0 a_r}^{LR}$ because we consider wavefunctions localized on the left, such that the correlation function has an exponential decay with respect to the first site ($0$). Similar results hold for $\Braket{a^\dagger_{L-1-r} a_{L-1}}^{RR}$ and $\Braket{a^\dagger_{L-1-r} a_{L-1}}^{RL}$ if ones calculate the correlation function between the last site of the lattice ($L-1$) and a site $r$ sites distant to the left. 
	
	\bigskip

\bibliography{non_herm}% Produces the bibliography via BibTeX.

\end{document}